\pdfoutput=1
\documentclass[aps,twocolumn,amsfonts,amssymb,prresearch,superscriptaddress,10pt,groupedaddress]{revtex4-2}

\usepackage{graphicx}
\graphicspath{{figures/}}
\usepackage{color}
\usepackage{xcolor}
\usepackage{amsfonts}
\usepackage{amsmath}
\usepackage{subfigure}
\usepackage[dvipsnames]{xcolor}
\usepackage[colorlinks=true,citecolor=blue]{hyperref}
\hypersetup{colorlinks=true,citecolor=blue,linkcolor=blue,urlcolor=blue}
\usepackage{outlines}
\usepackage{physics}
\usepackage{bm}
\usepackage{graphicx}
\usepackage{comment}

\allowdisplaybreaks

\begin{document}
\title{Viscous hydrodynamics from correlated losses}

\author{Aleksandr N. Mikheev}
\thanks{These authors contributed equally to this work.}
\author{Orjan Ameye}
\thanks{These authors contributed equally to this work.}
\author{Oded Zilberberg}
\author{Jeff Maki}
\email{jeffrey.maki@uni-konstanz.de}
\affiliation{Department of Physics, University of Konstanz, 78464, Konstanz, Germany}

\date{\today}

\begin{abstract}
Interactions turn a gas into a fluid. We show that two-body loss can take their place. Correlated loss induces elastic scattering, as required by the optical theorem, which dominates at strong loss via the quantum Zeno effect. A nondegenerate gas with only two-body loss thus flows hydrodynamically, with viscosity and thermal conductivity close to those of the unitary Fermi gas. As loss heats the gas, both grow in time with a universal exponent 3/8. Unlike single-particle loss, two-body loss can engineer collisions, and hence transport, in quantum matter.
\end{abstract}

\maketitle

\begin{figure}[t]
    \centering
    \includegraphics{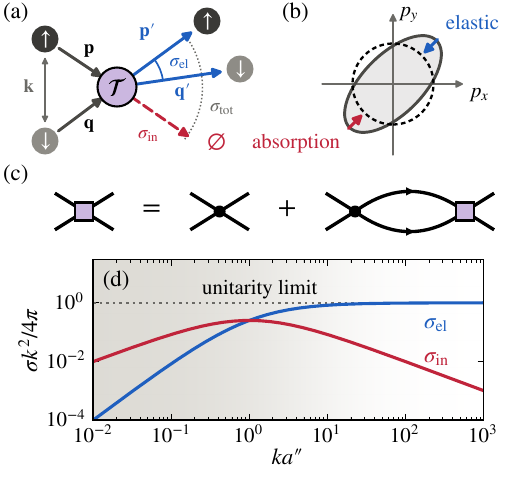}
    \caption{\textit{Loss-induced collisions.} (a) Two-body loss vertex \mbox{$\mathcal{T}$} depicting incoming particles with momenta $\bm{p}$ and $\bm{q}$ and relative momentum $k = |\bm{p}-\bm{q}|/2$ scattering via an elastic channel (solid arrows) into momenta $\bm{p}'$ and $\bm{q}'$ or an absorption channel (dashed arrow). (b) Relaxation of the distribution function \mbox{$f_{\sigma \bm{p}}$} (solid) toward local equilibrium (dashed) by elastic and absorption collisions. (c) Ladder equation for the two-body $T$-matrix of Eq.~\eqref{eq:Tmatrix} describing repeated two-particle scattering due to loss~\cite{SM}. (d) Elastic and absorption cross sections of a purely lossy two-body interaction, \mbox{$a=-ia''$}, in units of the unitary bound \mbox{$4\pi/k^2$}, versus \mbox{$ka''$} [cf.~Eq.~\eqref{eq:optical}]. Shading: \mbox{$\sigma_{\rm in}/\sigma_{\rm el}\sim\lambda_T/a''$} at the thermal momentum, \mbox{$ka''=\sqrt{2\pi}\,a''/\lambda_T$}, as in Fig.~\ref{fig:2}.}
    \label{fig:1}
\end{figure}


\textit{Introduction.--} Whether a gas flows like a fluid is decided by collisions between its constituents. In an ideal gas, particles move ballistically: each particle keeps its momentum, so currents of momentum density or heat persist until the particles reach the edge of the system. Collisions change this picture. They redistribute momentum and energy among the particles and bring the gas to local equilibrium, in which currents relax quickly to values set by the local density, velocity, and temperature. The gas then flows hydrodynamically, and momentum and heat spread diffusively at rates set by the shear viscosity and the thermal conductivity~\cite{LifshitzPitaevskii1981}. Nowadays, cold atoms can be used to realize both the collisionless and hydrodynamic regimes via careful control of the inter-atoms interactions~\cite{Chin2010}.  With the interactions switched off, the gas moves ballistically. In contrast, when the interactions are tuned to their maximum, each collision scatters with the largest probability that quantum mechanics allows, realizing the so-called unitary limit. In this limit, the gas flows hydrodynamically with one of the lowest viscosities measured in any fluid, as observed in two-component Fermi gases~\cite{Cao2011,Joseph2015,Enss2011}.

In experiment, ultracold gases are generally open systems that continuously lose particles from their trap. The combination of coherent motion with such lossy reactions is described by so-called quantum reaction-diffusion models~\cite{Sponselee2019,vanHorssen2015,Hofmann2022,Perfetto2023PRL,Perfetto2023PRE,Rowlands2024,Gerbino2024,Gerbino2025,Lehr2025,HM2026}. The simplest loss removes single particles independently of each other. It depletes the gas uniformly but leaves the motion of the remaining particles untouched. Hence, an ideal gas with single-particle loss stays ballistic. In contrast, two-body loss removes a particle only together with a partner it meets. It arises whenever two colliding particles leave the trap, e.g., when light binds two atoms into a molecule~\cite{Partridge2005,Tomita2017,Theis2004,Enomoto2008}, when atoms occupy metastable states~\cite{Tomita2019}, or when molecules react chemically~\cite{Syassen2008,Ospelkaus2010}. Since this loss requires two particles at the same place, its rate measures how likely particles are to be found close together~\cite{Werner2009,BHL2017,Gao2023}. As such, strong two-body loss even suppresses itself, because it acts as a continuous measurement of whether two particles overlap and thereby keeps them apart. This pertains to a quantum Zeno effect, which, for example, turns bosons in one dimension into impenetrable, hard-core particles~\cite{Syassen2008,GarciaRipoll2009,Zhu2014,Rossini2021,Rosso2022,Gerbino2024}.

Two-body loss thus enters ultracold atomic systems in two ways: its rate probes pair correlations, and, when strong, it keeps particles apart through the Zeno effect. In the settings studied so far, neither role relaxes the currents of the gas. Where loss drives a gas out of equilibrium and excites collective motion, the subsequent relaxation of hydrodynamic currents still commonly relies on collisions caused by the elastic interactions already present in the gas~\cite{Bouchoule2020,Maki2024}. Even the Zeno-induced hard-core repulsion produces no fluid in one dimension, as the system maps onto free fermions~\cite{Girardeau1960}. This raises the question of whether two-body loss alone, without a coherent elastic interaction, can turn a three-dimensional ideal gas into a fluid.

Here, we show that two-body loss alone can turn an ideal gas into a fluid, and that at strong loss this fluid is as collisional as the unitary Fermi gas, see Fig.~\ref{fig:1}. The mechanism follows from the optical theorem~\cite{LandauQM}: a process that removes colliding pairs must also scatter some of them elastically, see Fig.~\ref{fig:1}(a). Usually, this elastic part is a small correction to the loss. Strong loss, however, reverses the hierarchy. The Zeno effect keeps colliding particles apart, so pairs are rarely lost, while the elastic scattering induced by the loss saturates at the unitary limit, the largest value quantum mechanics allows. The gas then relaxes its currents far faster than it decays, see Fig.~\ref{fig:1}(b), and flows hydrodynamically while it slowly depletes.

To establish this, we derive a kinetic theory for a dilute three-dimensional Fermi gas from the Keldysh action of the lossy dynamics and solve it for the transport coefficients. Beyond the density decay studied in concurrent work~\cite{HM2026}, this yields the hydrodynamics of the decaying gas. Its shear viscosity and thermal conductivity approach those of the unitary Fermi gas, reached here without any coherent interaction. As the loss heats the gas, both grow in time with a universal exponent \mbox{$3/8$}, fixed by the unitary limit rather than by the loss strength. This growth provides a direct experimental signature of loss-driven hydrodynamics. Our results hold for bosons and fermions alike in the nondegenerate regime. More broadly, they show that loss, usually a nuisance that limits experiments, can serve as a resource: engineered dissipation~\cite{Diehl2008,Verstraete2009} can create the very collisions that make a fluid.

\textit{Model.--} We consider a homogeneous three-dimensional non-interacting Fermi gas subject to both single-particle and two-body loss in the non-degenerate regime. The density matrix evolves as
\begin{equation}
\partial_t \hat{\rho} = -i[\hat{H}, \hat{\rho}] + \int\dd[3]{\bm{x}}\Big(\gamma_1\sum_\sigma D_{\psi_{\sigma}} + 2\gamma_2 D_{\psi_{\downarrow}\psi_{\uparrow}}\Big)[\hat{\rho}],
\label{eq:Lindblad}
\end{equation}
with $\hbar=m=k_{\mathrm{B}}=1$. The coherent motion is governed by the free Hamiltonian
\begin{equation}
\hat{H}=\sum_\sigma\int \dd[3]{\bm{x}}\hat{\psi}^{\dagger}_{\sigma}(\bm{x})\left(-\frac{\nabla^2}{2}\right)\hat{\psi}_{\sigma}(\bm{x})\,,
    \label{eq:H}
\end{equation}
where $\hat{\psi}_{\sigma}(\bm{x})$ annihilates a particle with spin $\sigma\in\{\uparrow,\downarrow\}$ at position $\bm{x}$. The losses enter through Lindblad dissipators, $D_{L}[\hat{\rho}] = \hat{L}\hat{\rho} \hat{L}^{\dagger}- \tfrac{1}{2}\{\hat{L}^{\dagger}\hat{L},\hat{\rho}\}$, where the jump operator $\hat{L}$ specifies which particles are removed.

Single-particle loss, \mbox{$\hat{L}=\hat{\psi}_{\sigma}$}, removes particles with rate $\gamma_1$ and leaves the motion of the remaining particles untouched. Instead, two-body loss, $\hat{L}=\hat{\psi}_{\uparrow}\hat{\psi}_{\downarrow}$, removes two particles of opposite spin when they meet, at rate $2\gamma_2$~\cite{Huang2023,ZhangTajimaKato2026}. The anticommutator term in the dissipator adds \mbox{$i\gamma_2\int\dd[3]{\bm{x}}\hat{\psi}^{\dagger}_{\uparrow}\hat{\psi}^{\dagger}_{\downarrow}\hat{\psi}_{\uparrow}\hat{\psi}_{\downarrow}$} to the Hamiltonian, which is a contact interaction with imaginary coupling $i\gamma_2$. A coherent contact interaction of strength $g$, if present, would simply add to it, giving the complex bare coupling \mbox{$u=g+i\gamma_2$}~\cite{SM}.

The resulting elastic interactions can be seen by considering a pair of atoms with momenta $\bm{p}$ and $\bm{q}$. Due to Galilean invariance and rotational symmetry, the two-body scattering only depends on the magnitude of the relative momentum $k= |\bm{p}-\bm{q}|/2$, see Fig.~\ref{fig:1}(a). The two-body scattering is described by the two-body $T$-matrix, which accounts for repeated virtual collisions [cf.~Fig.~\ref{fig:1}(c)]
\begin{equation}
    \mathcal{T}(k)=\frac{4\pi}{a^{-1}+ik}\,.
    \label{eq:Tmatrix}
\end{equation}
Equation~\eqref{eq:Tmatrix} defines the complex $s$-wave scattering length $a=a'-ia''$, which is related to the bare coupling through \mbox{$u^{-1}=\Lambda/(2\pi^2)-1/(4\pi a)$}~\cite{Hutson2007,Idziaszek2010,BHL2017,SM}, where $\Lambda$ is the ultraviolet cutoff of the theory.

The probability for two-particles to scatter is related to the total cross section $\sigma_{\mathrm{tot}}$, which is related to the $T$-matrix according to the optical theorem
\begin{equation}
    -\mathrm{Im}\,\mathcal{T}(k) = k\,\sigma_{\mathrm{tot}}\,,\qquad \sigma_{\mathrm{tot}}=\sigma_{\mathrm{in}} +\sigma_{\mathrm{el}}\,.
    \label{eq:optical}
\end{equation}
The total cross section is composed of two parts, see Fig.~\ref{fig:1}(a). The absorption cross section \mbox{$\sigma_{\rm in}=|\mathcal{T}|^2\,\mathrm{Im}\,a^{-1}/(4\pi k)$} describes pairs ejected from the system and vanishes for $a''\to0$, so $a''$ controls the dissipation. The elastic cross section \mbox{$\sigma_{\rm el}=|\mathcal{T}|^2/(4\pi)$} redistributes the particles' momenta and depends on both $a'$ and $a''$. Without loss, $a''=0$, $\sigma_{\rm el}$ reduces to the standard s-wave result. Crucially, it remains finite even without intrinsic elastic scattering, $a'=0$.

For the remainder of this work, we will assume the system has no intrinsic elastic scattering \mbox{$a'=0$}, in order to study how two-body losses lead to emergent elastic two-body scattering. Hence, the ratio between the cross sections \mbox{$\sigma_{\rm el}/\sigma_{\rm in}$} denotes whether particles are mostly elastically scattered or removed by the two-particle loss, see Fig.~\ref{fig:1}(d). Using Eq.~\eqref{eq:Tmatrix}, this ratio is given by \mbox{$\sigma_{\rm el}/\sigma_{\rm in} = a'' k$}. For weak loss, \mbox{$a''k\ll1$}, pairs are mostly absorbed, at a rate per pair \mbox{$k\sigma_{\rm in}\simeq4\pi a''$} that grows with the loss strength, as in the textbook rate equation for two-body loss~\cite{Bouchoule2020}. For strong loss, \mbox{$a''k\gg1$}, the absorption rate instead decreases as \mbox{$k\sigma_{\rm in}\simeq 4\pi/(a''k)$}, while the elastic interactions approach their unitary value \mbox{$k\sigma_{\rm el} = 4\pi/k$}, leading to enhanced elastic collisions. For a thermal non-degenerate Fermi gas, scattering mostly occurs between atoms with relative momentum \mbox{$k\sim  1/\lambda_T$}, where \mbox{$\lambda_T = \sqrt{2\pi /T}$} is the thermal de Broglie wavelength. Hence, at high temperatures satisfying \mbox{$a''\gg\lambda_T$}, pairs of atoms counterintuitively scatter far more often than they are lost, and can collectively behave like a fluid.

The two resulting scattering channels become the collision terms of a Boltzmann equation. The latter governs the semiclassical distribution function $f_{\sigma{\bm{p}}} \equiv f_{\sigma}(t,{\bm{r}},{\bm{p}})$, i.e., the density of particles with spin \mbox{$\sigma$} and momentum $\bm{p}$ at position $\bm{r}$. Starting from the exact microscopic theory~\eqref{eq:Lindblad}, we derive the full Boltzmann using the two-particle-irreducible effective action formalism~\cite{Calzetta:2008iqa,Berges:2015kfa,SM}
\begin{equation}
    \left(\partial_t+{\bm{p}}\cdot\nabla_{\bm{r}}+\gamma_1\right)f_{\sigma\bm{p}}=I^{\rm in}[f_{\sigma \bm{p}}]+ I^{\rm el}[f_{\sigma\bm{p}}]\,.
    \label{eq:BE}
\end{equation}
The left-hand side describes free streaming, i.e., each particle moving along a straight line with velocity $\bm{p}$, while single-particle loss enters only as a uniform decay. Since this decay leaves the motion of the remaining particles untouched, we set $\gamma_1=0$ from now on. The right-hand side contains the two channels of Eq.~\eqref{eq:optical}. Absorption removes pairs,
\begin{equation}
    I^{\rm in}[f_{\sigma \bm{p}}]=-\frac{\mathrm{Im}\,a^{-1}}{2\pi}\int_{\bm{q}}\left|\mathcal{T}(k)\right|^2 f_{\sigma\bm{p}}f_{\bar{\sigma} \bm{q}}\,,
    \label{eq:Inelastic}
\end{equation}
where $\int_{\bm{q}} = \int\dd[3]{\bm{q}}/(2\pi)^3$ and $k= |\bm{p}-\bm{q}|/2$ is again the magnitude of the relative momentum. Elastic scattering redistributes momentum,
\begin{equation}
    I^{\rm el}[f_{\sigma\bm{p}}]=\int_{\bm{q}\bm{p'}\bm{q'}}\mathcal{W}_{\bm{p}\bm{q}; \bm{p'}\bm{q'}}\left[f_{\sigma\bm{p'}}f_{\bar{\sigma}\bm{q'}}-f_{\sigma\bm{p}}f_{\bar{\sigma}\bm{q}}\right]\,,
    \label{eq:Elastic}
\end{equation}
with the transition rate
\begin{align}
    \mathcal{W}_{\bm{p}\bm{q}; \bm{p}'\bm{q}'} 
    &=
    \left|\mathcal{T}(k)\right|^2 (2\pi)^4\delta^{(3)}(\bm{p}+\bm{q}-\bm{p}'-\bm{q}') \nonumber \\
    &\times\delta(\epsilon_{\bm p}+\epsilon_{\bm q}-\epsilon_{\bm p'}-\epsilon_{\bm q'})\,,
    \label{eq:rate}
\end{align}
where $\epsilon_{\bm p}=\bm{p}^2/2$. Crucially, Eq.~\eqref{eq:Elastic} is identical to the collision integral of a gas with real scattering length \mbox{$a=a'$}. Here, however, it arises entirely from the two-body loss. Equations~\eqref{eq:BE}\,--\,\eqref{eq:rate} are a central result of this work. They are valid in the non-degenerate regime where \mbox{$f_{\sigma\bm{p}}\ll1$}. In this limit, effects from particle statistics are irrelevant, and our results hold also for bosonic particles~\cite{SM}. Note that a kinetic equation of this form was also obtained independently in concurrent work focusing on density decay in arbitrary dimensions~\cite{HM2026}.

\begin{figure}[t]
    \centering
    \includegraphics{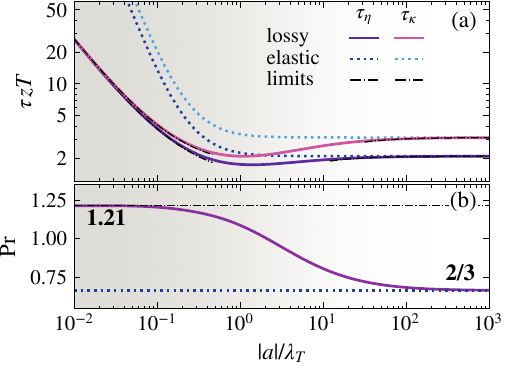}
    \caption{\textit{Instantaneous relaxation times.} (a) Shear and thermal relaxation times, in units of \mbox{$1/(zT)$}, versus \mbox{$|a|/\lambda_T$} for loss, \mbox{$a=-ia''$} (solid; shear purple, heat magenta), and for conventional elastic scattering, \mbox{$a=a'$} (dotted; shear dark blue, heat light blue). Black dash-dotted lines: asymptotes of Eqs.~\eqref{eq:tau_strong} and \eqref{eq:tau_weak}. Shading: ratio of the absorption and elastic cross sections, \mbox{$\sigma_{\rm in}/\sigma_{\rm el}\sim\lambda_T/a''$}, at the thermal momentum, graded from white (\mbox{$\ll1$}, hydrodynamic) to grey (\mbox{$\gg1$}, loss-dominated). (b) Prandtl number, from the universal value \mbox{$2/3$} (dotted) at \mbox{$|a|\gg\lambda_T$} to the steady-distortion value \mbox{$\approx1.21$} (dash-dotted) at \mbox{$|a|\ll\lambda_T$}.}
    \label{fig:2}
\end{figure}

\textit{Analysis of the Boltzmann equation.--} We now analyze the dynamics of a gas subject to Eq.~\eqref{eq:BE}. We  start from a Maxwell--Boltzmann (MB)  distribution function \mbox{$f_{\sigma \bm{p}}(0)= (n_0\lambda_{T_0}/2)\,\mathrm{e}^{-p^2/(2T_0)}$} describing a homogeneous gas \mbox{$n_0$} at temperature \mbox{$T_0$}. We quantify the hydrodynamic flow by examining the relaxation of the distribution function towards local equilibrium~\cite{LifshitzPitaevskii1981}, see Fig.~\ref{fig:1}(b). Specifically, we extend the Chapman--Enskog expansion~\cite{ChapmanCowling1970,LifshitzPitaevskii1981,SmithJensen1989} to lossy gases, and expand the full distribution around a time-dependent distribution \mbox{$f = f^{(0)}_{\sigma \bm{p}} + \delta f_{\sigma \bm{p}}$}, and linearize the Boltzmann equation. The expansion parameter is the Knudsen number, which is the ratio of the mean free path, i.e., the distance a particle travels between elastic collisions, to the length over which the density, velocity, and temperature vary.

The zeroth order solution  \mbox{$f^{(0)}_{\sigma \bm{p}}$} describes the behavior of the time-dependent local equilibrium, which is not necessarily a MB distribution due to the losses. It obeys the spatially homogeneous Boltzmann equation, 
\begin{equation}
\partial_t f^{(0)}_{\sigma{\bm p}}=I^{\rm in}[f^{(0)}_{\sigma{\bm p}}]+I^{\rm el}[f^{(0)}_{\sigma{\bm p}}]\,
\end{equation}
 which we solve by writing \mbox{$f^{(0)}_{\sigma{\bm p}}$} as a MB distribution times \mbox{$1+\psi_{\bm p}$}. The small distortion \mbox{$\psi_{\bm p}$} does not contribute to the density or energy and can be exactly evaluated by expanding the zeroth order Boltzmann equation in orthogonal polynomials~\cite{SM}. Thus, by construction, the time-dependence of the density and temperature are encoded in the MB distribution, which we extract by computing the moments of the distribution function: \mbox{$n(t)=\sum_{\sigma} \int_{\bm p} f_{\sigma \bm{p}}^{(0)}(t)$} and \mbox{$\mathcal{E}(t)=\sum_{\sigma} \int_{\bm p} \epsilon_{\bm{p}} f_{\sigma \bm{p}}^{(0)}(t) = \frac{3}{2}n(t) T(t)$}, for the number and energy density, respectively. 

Using this procedure, we find for weak two-body loss, \mbox{$a''\ll\lambda_T$}, that all momenta are lost at the same rate. The gas then decays according to the textbook rate equation, \mbox{$\dot n=-4\pi a''n^2$}, i.e., \mbox{$n\propto t^{-1}$}, while the temperature is approximately constant. In this case, the lack of elastic interactions and non-equilibrium nature of the losses lead to a departure from the MB distribution function~\cite{SM}. Strong loss, \mbox{$a''\gg\lambda_T$}, instead, changes both observations. At thermal momenta, $k\sim  1/\lambda_T$, a pair is now about \mbox{$a''/\lambda_T$} times more likely to scatter than to be lost, independent of the density, cf.~the shading in Fig.~\ref{fig:2}. Hence, the gas quickly approaches a local equilibrium state that is approximately a MB distribution, with a density \mbox{$n(t)$} and temperature \mbox{$T(t)$} that change slowly as the gas decays, up to small corrections of order \mbox{$\mathcal{O}(\lambda_T/a'')$}. Since the absorption rate per pair decreases with relative momentum, slow pairs are lost preferentially, and the remaining gas heats up. This leads to the following universal scaling behavior for the temperature and density~\cite{SM}
\begin{equation}
T\propto t^{1/4},\qquad n\propto t^{-3/4}.
\label{eq:decay}
\end{equation}
As \mbox{$\lambda_T$} shrinks with heating, \mbox{$a''/\lambda_T$} grows in time, and the gas becomes ever more collisional. 

The first order correction to the distribution functions, \mbox{$\delta f$}, describes deformations that produce shear stress and heat currents and satisfy the linearized Boltzmann equation
\begin{equation}
\partial_t\,\delta f_{\sigma \bm p}-I'_{\bm p}[\delta f]=-f^{(0)}_{\sigma \bm p}H_{\bm p},
\label{eq:first_main_text}
\end{equation}
where \mbox{$H_{\bm p}$} are sources of the deformations due to external shear flows or temperature gradients, while \mbox{$I'_{\bm p}[\delta f]$} is the linearized collision integral~\cite{SM}. These deformations relax due to elastic and absorption collisions on time scales \mbox{$\tau_\eta$} and \mbox{$\tau_\kappa$}~\cite{LifshitzPitaevskii1981}, see Fig.~\ref{fig:1}(b). We calculate these relaxation times using a single mode expansion of the linearized Boltzmann equation~\cite{Hofmann2023, MakiEnss2023, SM}, which captures the slowest relaxation channel of the collision integral. The relaxation times set the shear viscosity \mbox{$\eta=nT\tau_\eta$} and the thermal conductivity \mbox{$\kappa=\tfrac52nTr_T\tau_\kappa$}. Since we have set \mbox{$a'=0$}, this relaxation is entirely due to two-body losses. We note that losses also lead to a non-trivial correction to the thermal conductivity proportional to {$r_T=1+\mathcal O(\lambda_T/a'')$}. This term originates from the nonequilibium nature of the losses which preferentially removes slow momenta~\cite{SM}.

For \mbox{$a''\gg\lambda_T$}, where hydrodynamics holds, both relaxation times approach their unitary values, see Fig.~\ref{fig:2}(a),
\begin{equation}
\tau_{\eta(\kappa)}=\frac{d_{\eta(\kappa)}\tau_u}{zT}\left(1-c_{\eta (\kappa)}\frac{\lambda_T}{a''}\right)\,,
\label{eq:tau_strong}
\end{equation}
with fugacity \mbox{$z$}, \mbox{$n=2z\lambda_T^{-3}$}, \mbox{$\lambda_T=\sqrt{2\pi/T}$}, the unitary constant \mbox{$\tau_u=15\pi/(16\sqrt2)$}, \mbox{$d_{\eta(\kappa)}=1\ (3/2)$}, and \mbox{$c_{\eta(\kappa)}=1.208\ (2.964)$}. The leading term is the relaxation time of the unitary Fermi gas~\cite{Massignan2005,BruunSmith2005,Enss2011,Hofmann2020,Frank2020,Fujii2021,Joseph2015,Baird2019}, reached here without any coherent interaction, and coincides with the elastic result for \mbox{$|a|\gg\lambda_T$}, cf.~the dotted lines in Fig.~\ref{fig:2}(a). The loss correction shortens both relaxation times, but it reduces the thermal relaxation time more due to the distortion of the local-equilibrium distribution function.

For \mbox{$a''\ll\lambda_T$}, absorption dominates. The shear and thermal deformations then relax at the rate at which the gas itself disappears, so the coefficients in this regime do not describe a hydrodynamic flow, cf.~the dark shading in Fig.~\ref{fig:2}. Notwithstanding, the relaxation times remain well defined
\begin{equation}
\tau_{\eta(\kappa)}=\frac{1}{zT}\left(\frac{\lambda_T}{4a''}+C_{\eta(\kappa)}\right),
\label{eq:tau_weak}
\end{equation}
with \mbox{$C_{\eta}=9\sqrt2/10$} and \mbox{$C_{\kappa}=137\sqrt2/120$}~\cite{SM}. At leading order in \mbox{$a''/\lambda_T$}, shear and heat relax at the common loss rate. 

We quantify the relative strength of momentum and thermal diffusivity using the Prandtl number. Although this is defined in equilibrium, the frequent elastic interactions in our system give a well defined time-dependent Prandtl number \mbox{$\mathrm{Pr}=c_p\eta/\kappa=\tau_\eta/(r_T\tau_\kappa)$}, with \mbox{$c_p=5/2$} being the heat capacity at constant pressure. Fig.~\ref{fig:2}(b) shows the Prandtl number as a function of \mbox{$a''/\lambda_T$}. It takes the universal value \mbox{$2/3$} for \mbox{$a''\gg\lambda_T$}~\cite{Frank2020,Fujii2021}. For \mbox{$a''\ll\lambda_T$}, where \mbox{$\tau_\eta\simeq\tau_\kappa$}, it rises to \mbox{$\mathrm{Pr}=1/r_T\approx1.21$}, the value for a steady distortion. In a decaying gas the distortion builds up slowly, and \mbox{$\rm Pr$} grows in time toward this value, reaching \mbox{$\mathrm{Pr}\approx1.15$} at \mbox{$a''\approx0.3\lambda_T$}, well above the value \mbox{$2/3$} of a dilute gas.

\begin{figure}[t]
    \centering
    \includegraphics{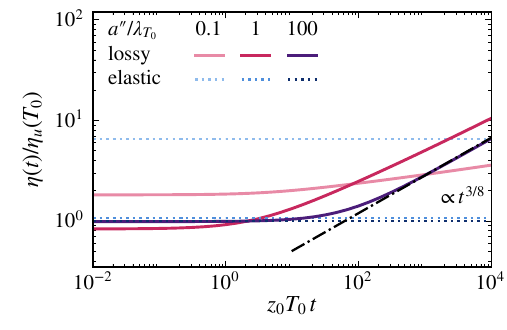}
    \caption{\textit{Viscosity along the decay.} Shear viscosity in units of the unitary value \mbox{$\eta_u(T_0)=2\tau_u\lambda_{T_0}^{-3}$} versus \mbox{$z_0T_0t$}, for initial \mbox{$a''/\lambda_{T_0} = 0.1,\ 1,\ 100$} (solid) and for elastic scattering at equal \mbox{$|a|/\lambda_{T_0}$} and fixed \mbox{$T_0$} (dotted). Dash-dotted: long-time power law \mbox{$\eta\propto t^{3/8}$} under loss.}
    \label{fig:3}
\end{figure}

So far, we have only discussed the instantaneous relaxation rates, i.e., the rates at a given \mbox{$a''/\lambda_T$}. However, since the density and temperature are time-dependent the shear viscosity and thermal conductivity also evolve with time. In  Fig.~\ref{fig:3}, we  show the time-dependence of the shear viscosity, for fixed initial \mbox{$a''/\lambda_{T_0}$}. For \mbox{$a''\gg\lambda_{T_0}$}, Eq.~\eqref{eq:tau_strong} gives \mbox{$\eta\simeq2\tau_u\lambda_T^{-3}\propto T^{3/2}$} and similarly \mbox{$\kappa \propto T^{3/2}$}. We emphasize that this result is universal and is set by the nearby unitary point \mbox{$1/a=0$}. 
In contrast for \mbox{$a''\ll\lambda_T$}, \mbox{$\eta$} and \mbox{$\kappa$} grow only slowly. However, as the gas heats up, the thermal de Broglie wavelength shrinks, so the gas will approach the universal scaling law with exponent \mbox{$3/8$} when  \mbox{$a''/\lambda_T \gg 1$}.

\textit{Discussion.--} We have shown that two-body loss counterintuitively turns an ideal nondegenerate quantum gas into a near-unitary fluid. The fluid owes its existence to the optical theorem as colliding pairs must lead to both inelastic and elastic scattering. Single-particle loss, by contrast, removes particles individually and thus cannot generate elastic collisions. Our analysis holds for a nondegenerate gas at intermediate times, while the density remains appreciable. At low temperature, quantum statistics and the many-body background add terms to Eq.~\eqref{eq:BE} that enhance the dissipation~\cite{SM}. In this limit, other loss channels may dominate, as in Bose gases where there is strong three-body recombination~\cite{Burt1997,Weber2003,Pollack2009,Eismann2016}.

Cold atoms offer two routes to test these predictions. Near the broad Feshbach resonance of \mbox{$^6$Li} at \mbox{$832.18\,\mathrm{G}$}~\cite{Zurn2013}, the gas is already a unitary fluid, and photoassociation, previously used to probe the closed-channel fraction and thereby the contact~\cite{Partridge2005,Werner2009}, adds a tunable \mbox{$a''$}. Because our kinetic theory holds for general complex \mbox{$a$}, it predicts a loss-induced reduction of both relaxation times in this regime, observable in the damping of sound modes in box traps~\cite{Cao2011,Joseph2015}. A direct test of loss-induced fluidity, where loss alone makes the gas hydrodynamic, instead requires \mbox{$a'\approx0$}, e.g., near the zero crossing of the scattering length at \mbox{$527\,\mathrm{G}$}~\cite{Zurn2013}, together with \mbox{$a''\gg\lambda_T$}. This calls for a resonantly enhanced loss channel. Photoassociation in \mbox{$^7$Li} has reached the unitarity limit of two-body loss, \mbox{$ka''\sim1$}~\cite{Junker2008}. A Feshbach resonance with a decaying closed channel is also viable, as coupled-channel calculations for \mbox{$^{85}$Rb} near \mbox{$155\,\mathrm{G}$}  predict a peak \mbox{$a''\approx1.1\times10^5a_0\approx6\,\mu\mathrm{m}$}~\cite{Hutson2007}, i.e., \mbox{$a''/\lambda_T\approx30$}--\mbox{$55$} at \mbox{$1$}--\mbox{$3\,\mu\mathrm{K}$}.

Our results rest on the optical theorem and generally hold for any absorbing few-body process. Every correlated dissipator then comes with a coherent scattering channel, and strong dissipation can potentially make this channel dominate. Engineered loss thus offers a route to engineered collisions, and hence transport, in quantum matter. This framework naturally extends to other open quantum many-body systems, such as single-component \mbox{$p$}-wave molecular gases, where loss is intrinsic~\cite{Gao2023,GaoYan2025PRL}, driven-dissipative fluids of light, where gain joins loss~\cite{CarusottoCiuti2013}, and the degenerate regime, where quantum statistics modifies the loss-induced collisions.

\textit{Acknowledgements.--} We acknowledge funding from the Deutsche Forschungsgemeinschaft (DFG) via project numbers 449653034 (Heisenberg), 425217212 (SFB1432), 521530974 (FOR5688), from the German-Israeli Foundation for Scientific Research and Development (GIF, Grant No. G-3012-303.10/2026), as well as from the Swiss National Science Foundation (SNSF) through NCCR SPIN.

\textit{Note added.--} While preparing this manuscript, we became aware of Ref.~\cite{HM2026}, which derives an equivalent Boltzmann equation and focuses on the density and temperature dynamics in arbitrary dimension. Our work is complementary and addresses the induced hydrodynamic transport generated by the loss-induced collisions. We differ in regards to the universal scaling law for large imaginary scattering lengths, which is due to Ref.~\cite{HM2026} having a small real scattering length set by the ultraviolet cutoff of the theory.


\newpage
\begin{widetext}
\begin{center}
\textbf{Supplementary Material}
\end{center}
\setcounter{equation}{0}
\setcounter{figure}{0}
\setcounter{table}{0}
\setcounter{section}{0}
\renewcommand{\theequation}{S\arabic{equation}}
\renewcommand{\thefigure}{S\arabic{figure}}
\renewcommand{\thetable}{S\Roman{table}}

In the supplementary material, we provide further details of the theoretical frameworks employed in this work and results.

\section{Derivation of the Boltzmann equation}
\subsection{Keldysh formalism}

For open quantum systems whose evolution is governed by the Lindblad master equation, the Keldysh action $d$ spatial dimensions is given by~\cite{Sieberer2016,Sieberer:2023qyv}
\begin{equation}
\label{eq:S}
S 
=
\int\dd{t}\dd[d]{\bm{x}} \left\{\vphantom{\frac{1}{2}}i\bar{\psi}^{+}_{\sigma}\partial_t\psi^{+}_{\sigma} - H(\psi^{+},\bar{\psi}^{+})  - i\bar{\psi}_{\sigma}^{-}\partial_t\psi_{\sigma}^{-} + H(\psi^{-},\bar{\psi}^{-}) 
-
i \sum_{\alpha} \gamma_{\alpha} \left[L_{\alpha}^{+} \bar{L}_{\alpha}^{-} - \frac{1}{2}\left(\bar{L}_{\alpha}^{+} L_{\alpha}^{+} + \bar{L}_{\alpha}^{-} L_{\alpha}^{-}\right)\right]\right\},
\end{equation}
where $\pm$ denotes the forward and backward branches of the Keldysh time-countour, $H_{\pm} = H(\bar{\psi}^{\pm},\psi^{\pm})$, $L^{+}=L(\bar{\psi}^{+},\psi^{+})$, $L^{-} = L(\zeta\bar{\psi}^{-},\zeta\psi^{-})$, $\bar{L}^{+}=L^\dagger(\bar{\psi}^{+},\psi^{+})$, and $\bar{L}^{-} = L^{\dagger}(\zeta\bar{\psi}^{-},\zeta\psi^{-})$. These functions follow from $\hat{H}$, $\hat{L}$, and $\hat{L}^{\dagger}$, which are assumed to be normally ordered, by substituting $\bar{\psi}$ ($\psi$) for creation (annihilation) operators. The factor $\zeta=1$ ($-1$) for bosons (fermions) in $L^{(\dagger)}_-$ flips the sign of the Grassmann fields on the backward branch. 

The first part of the Keldysh action \eqref{eq:S} represents the unitary part of the dynamics, governed by the Hamiltonian density $H$, while the last term accounts for the dissipative part of the Lindblad dynamics. Although in this work we focus on a non-interacting case, for generality, in this supplemental material we will also include an attractive $s$-wave density-density contact interaction:
\begin{equation}
\hat{H} = \int\dd[d]{\bm{x}} \left[\hat{\psi}^{\dagger}_{\sigma}(\bm{x})\left(-\nabla_{\bm{x}}^2/2 + U(\bm{x})\right) \hat{\psi}_{\sigma}(\bm{x}) + g\,\hat{\psi}^{\dagger}_{\uparrow}(\bm{x}) \hat{\psi}^{\dagger}_{\downarrow}(\bm{x}) \hat{\psi}_{\uparrow}(\bm{x}) \hat{\psi}_{\downarrow}(\bm{x}) \right],
\end{equation}
where $U(\bm{x})$ is a external potential that we keep for full generality and $g$ is the interaction strength. Combined with the dissipation channels, given by single-fermion losses $\hat{L}_{\sigma}(\bm{x}) = \hat{\psi}_{\sigma}(\bm{x})$ at rate $\gamma_1$ and local two-fermion losses $\hat{L}(\bm{x}) = \hat{\psi}_{\uparrow}(\bm{x})\,\hat{\psi}_{\downarrow}(\bm{x})$ at rate $2 \gamma_2$, the resulting Keldysh action reads
\begin{equation}
S = \int\dd{t}\dd[d]{\bm{x}} \left(\bar{\psi}^{+}_{\sigma} \enspace \bar{\psi}_{\sigma}^{-}\right) 
\begin{pmatrix}
i\partial_t + \frac{1}{2}\nabla_{\bm{x}}^2 - U(\bm{x}) + \frac{i}{2} \gamma_1 & 0 \\
-i\gamma_1 & -\left(i\partial_t + \frac{1}{2}\nabla_{\bm{x}}^2 - U(\bm{x})\right) + \frac{i}{2} \gamma_1
\end{pmatrix}
\begin{pmatrix}
\psi^{+}_{\sigma} \\
\psi^{-}_{\sigma}
\end{pmatrix} 
+
S_{\mathrm{int}}\,,
\end{equation}
where the interaction term reads
\begin{align}
S_{\mathrm{int}} 
&=
-\int\dd{t}\dd[d]{\bm{x}} \left\lbrace g \left[\psi^{+}_{\uparrow} \psi^{+}_{\downarrow} \bar{\psi}^{+}_{\uparrow}\bar{\psi}^{+}_{\downarrow} - \psi^{-}_{\uparrow} \psi^{-}_{\downarrow} \bar{\psi}^{-}_{\uparrow}\bar{\psi}^{-}_{\downarrow} \right] + 2 i\gamma_2 \left[\psi^{+}_{\uparrow} \psi^{+}_{\downarrow} \bar{\psi}^{-}_{\downarrow}\bar{\psi}^{-}_{\uparrow} - \frac{1}{2}\left(\bar{\psi}^{+}_{\downarrow} \bar{\psi}^{+}_{\uparrow} \psi^{+}_{\uparrow}\psi^{+}_{\downarrow} + \bar{\psi}^{-}_{\downarrow} \bar{\psi}^{-}_{\uparrow} \psi^{-}_{\uparrow}\psi^{-}_{\downarrow}\right) \right]\right\rbrace\nonumber\\
&\equiv
-\int\dd{t}\dd[d]{\bm{x}}\,J^{\dagger} \cdot K \cdot J\,,
\end{align}
with 
\begin{equation}
K 
= 
\begin{pmatrix}
u & 0  \\
u - \bar{u} & -\bar{u} \\
\end{pmatrix},\qquad  
J
= 
\begin{pmatrix}
\phantom{-}\psi^{+}_{\uparrow} \psi^{+}_{\downarrow}\\
-\psi^{-}_{\uparrow}\psi^{-}_{\downarrow}\\
\end{pmatrix},\qquad
J^{\dagger}
= 
\left(\bar{\psi}^{+}_{\uparrow} \bar{\psi}^{+}_{\downarrow},
-\bar{\psi}^{-}_{\uparrow}\bar{\psi}^{-}_{\downarrow}\right),
\end{equation}
and we have introduced a complex coupling $u \equiv g + i\gamma_2$ and its complex conjugate $\bar{u} \equiv g - i\gamma_2$. We decouple this interaction in the pairing channel as
\begin{equation}
S_{\mathrm{int}}
\to
-\int\dd{t}\dd[d]{\bm{x}} \left(\phi^{\dagger}\cdot K^{-1} \cdot \chi + i J^{\dagger} \cdot \chi + i\phi^{\dagger} \cdot J\right), 
\end{equation}
where $\chi = (\chi^{+},\chi^{-})^T$ and $\phi^{\dagger} = (\bar{\phi}^{+},\bar{\phi}^{-})$ are complex-valued Keldysh bilinears. Note that we have introduced two independent complex Hubbard--Stratonovich fields, $\phi$ and $\chi$, to make explicit that the fields coupling to the pair and conjugate-pair operators are independent integration variables in the non-Hermitian Keldysh theory, rather than complex conjugates of each other. One could equivalently denote them by $\phi$ and $\bar{\phi}$, provided the bar is understood merely as a label for an independent field and not as complex conjugation; using distinct symbols avoids this potential ambiguity.

Upon decoupling the interaction, the action takes the form $S = S_{\mathrm{F}} + S_{\mathrm{HS}} + S_{\mathrm{int}}$, where $S_{\mathrm{F}}$ and $S_{\mathrm{HS}}$ denote the quadratic fermionic and Hubbard--Stratonovich parts, respectively, and $S_{\mathrm{int}}$ denotes the interaction part. Moving to the Keldysh basis,  
\begin{equation}
O^{\mathrm{cl}} \equiv \frac{O^{+} + O^{-}}{2}\,,\qquad O^{\mathrm{q}} \equiv O^{+} - O^{-}\,,
\end{equation}
we can write the quadratic part of the action as 
\begin{align}
\label{eq:S0}
S_{\mathrm{F}} 
&=
\int\dd{t}\dd[d]{\bm{x}} \left(\bar{\psi}^{\mathrm{cl}}_{\sigma} \enspace \bar{\psi}_{\sigma}^{\mathrm{q}}\right) 
\begin{pmatrix}
0 & i\partial_t + \frac{1}{2}\nabla_{\bm{x}}^2 - U(\bm{x}) -\frac{i}{2}\gamma_1 \\
i\partial_t + \frac{1}{2}\nabla_{\bm{x}}^2 - U(\bm{x}) + \frac{i}{2}\gamma_1 & \frac{i}{2}\gamma_1
\end{pmatrix}
\begin{pmatrix}
\psi_{\sigma}^{\mathrm{cl}} \\
\psi_{\sigma}^{\mathrm{q}}
\end{pmatrix},\nonumber\\
S_{\mathrm{HS}}
&=
-\int\dd{t}\dd[d]{\bm{x}}
\left(\bar{\phi}^{\mathrm{cl}}\enspace \bar{\phi}^{\mathrm{q}}\right) 
\begin{pmatrix}
0 & 1/\bar{u} \\
1/u & -i\gamma_2/|u|^2
\end{pmatrix}
\begin{pmatrix}
\chi^{\mathrm{cl}} \\
\chi^{\mathrm{q}}
\end{pmatrix},
\end{align}
while the interaction term takes the form
\begin{align}
S_{\mathrm{int}}
&=
-i\int\dd{t}\dd[d]{\bm{x}}\left[
\chi^{\mathrm{q}}\left(\bar\psi^{\mathrm{cl}}_{\uparrow}\bar{\psi}^{\mathrm{cl}}_{\downarrow} + \frac{1}{4}\bar{\psi}^{\mathrm{q}}_{\uparrow}\bar{\psi}^{\mathrm{q}}_{\downarrow}\right)
+
\chi^{\mathrm{cl}}\left(\bar{\psi}^{\mathrm{cl}}_{\uparrow}\bar{\psi}^{\mathrm{q}}_{\downarrow} +\bar{\psi}^{\mathrm{q}}_{\uparrow}\bar{\psi}^{\mathrm{cl}}_{\downarrow}\right)
+ 
\left(\psi^{\mathrm{cl}}_{\uparrow}\psi^{\mathrm{cl}}_{\downarrow} +\frac{1}{4}\psi^{\mathrm{q}}_{\uparrow}\psi^{\mathrm{q}}_{\downarrow}\right)\bar{\phi}^{\mathrm{q}}
+
\left(\psi^{\mathrm{cl}}_{\uparrow}\psi^{\mathrm{q}}_{\downarrow} +\psi^{\mathrm{q}}_{\uparrow}\psi^{\mathrm{cl}}_{\downarrow}\right) \bar{\phi}^{\mathrm{cl}}
\right]\nonumber\\
&\equiv
-i\int\dd{t}\dd[d]{\bm{x}} \gamma_{abc}\left(\chi^a \bar{\psi}^{b}_{\uparrow}\bar{\psi}^{c}_{\downarrow} + \psi_{\uparrow}^a \psi_{\downarrow}^b \bar{\phi}^c\right),
\end{align}
where $\gamma_{abc}$ is a totally symmetric tensor whose only nonvanishing components are $\gamma_{\mathrm{clclq}} = \gamma_{\mathrm{clqcl}} = \gamma_{\mathrm{qclcl}} = 1$ and $\gamma_{\mathrm{qqq}} = 1/4$. Note that, following Larkin and Ovchinnikov~\cite{Kamenev2011}, it is customary in the literature to use transformations with opposite signs for $\psi$ and $\bar{\psi}$ and not to denote the rotated fields as ``classical and quantum'', since fermionic fields do not have a classical interpretation. In this work, we depart from both conventions in order to maintain uniform notation for fermions and composite Hubbard--Stratonovich bosons.

\subsection{Kadanoff--Baym equations}
Our derivation of the Boltzmann equation featured in the main text starts from a set of self-consistent equations for the one- and two-point correlation functions. The latter are the natural objects for kinetic theory, since the dressed propagators carry information on both the spectral properties of the particles and their distribution in phase space. Closed equations for such correlation functions can be conveniently obtained from a variational principle for the so-called two-particle (2PI) irreducible effective action, which treats mean fields and propagators on an equal footing ~\cite{Berges:2004yj,Calzetta:2008iqa,Rammer:2007,Berges:2015kfa}.
The 2PI effective action can be written in the standard form as~\cite{Berges:2004yj,Rammer:2007,Calzetta:2008iqa,Berges:2015kfa}
\begin{equation}
\label{eq:Gamma_2PI}
\Gamma_{\mathrm{2PI}} = S - \frac{i}{2} \Tr\ln G^{-1} + \frac{i}{2} \Tr\ln D^{-1} - \frac{i}{2} \Tr \left(G_0^{-1}\,G\right) + \frac{i}{2}
\Tr \left(D_0^{-1}\,D\right) + \Gamma_2\,.
\end{equation}
Here, $G$ ($G_0$) and $D$ ($D_0$) are the dressed (bare) propagators for the fermionic and the Hubbard--Stratonovich fields, respectively, and  $\Gamma_2 = -i\ln\langle\exp(i S_{\mathrm{int}})\rangle^{\mathrm{2PI}}_{G,D}$ is given by the sum of all connected vacuum diagrams that are two-particle irreducible with respect to the full propagators $G$ and $D$. Note that the prefactor $1/2$, absent in the standard complex-fermion formulation \cite{Bock:2015evx}, arises from the Nambu doubling introduced below, $\Psi=(\psi,\bar{\psi})^T$, which counts each physical degree of freedom twice \cite{Ruester:2004eg}.

Introducing a compact superfield notation,
\begin{equation}
\Psi_{1\sigma}^a = \psi_{\sigma}^a,\quad \Psi_{2\sigma}^a = \bar{\psi}_{\sigma}^a,\quad \Phi_{11}^a = \phi^a,\quad \Phi_{21}^a = \bar{\phi}^a,\quad  \Phi_{12}^a = \chi^a,\quad \Phi_{22}^a = \bar{\chi}^a,
\end{equation}
where $a \in \{\mathrm{cl},\mathrm{q}\}$, the propagators are defined using the standard convention
\begin{equation}
D_{ij;\alpha\beta}^{ab}(x,x') = \langle\Phi_{i\alpha}^a(x)\,\bar{\Phi}_{j\beta}^b(x')\rangle_c\,,\qquad
G_{ij;\sigma\sigma'}^{ab}(x,x') = \langle\Psi_{i\sigma}^a(x)\,\bar{\Psi}_{j\sigma'}^b(x')\rangle,
\end{equation}
 while the inverse bare propagators are given by
\begin{equation}
i\left(D_0^{-1}\right)^{ab}_{ij;\alpha\beta}(x,x') = \frac{\delta^2 S[\Psi,\Phi]}{\delta \bar{\Phi}^a_{i\alpha}(x)\,\delta\Phi^b_{j\beta}(x')}\,,\qquad 
i\left(G_0^{-1}\right)_{ij;\sigma\sigma'}^{ab}(x,x')
=
\frac{\overrightarrow{\delta}}{\delta \bar{\Psi}^a_{i\sigma}(x)} S[\Psi,\Phi] \frac{\overleftarrow{\delta}}{\delta \Psi^b_{j\sigma'}(x')},
\end{equation}
where all the derivatives are evaluated at the field expectation values. For the considered theory, the only nonvanishing Hubbard--Stratonovich entries are
\begin{equation}
i\left(D_0^{-1}\right)^{ab}_{11;12}(x,x') 
=
-
\begin{pmatrix}
0 & 1/\bar{u} \\
1/u & -i\gamma_2/|u|^2
\end{pmatrix}
\delta(x-x')\,,\qquad
i \left(D_0^{-1}\right)^{ab}_{22;21}(x,x') 
=
-
\begin{pmatrix}
0 & 1/u \\
1/\bar{u} & -i\gamma_2/|u|^2
\end{pmatrix}
\delta(x-x')\,,
\end{equation}
while the fermionic inverse bare propagators take the form
\begin{align}
\label{eq:G0_inv}
i\left(G_0^{-1}\right)_{ij;\sigma\sigma'}^{ab}(x,x')
&=
\delta(x-x') \left\lbrace\delta_{\sigma\sigma'}\left[\sigma_x^{ab}\left(i\delta_{ij} \partial_{t'} + \sigma^z_{ij}\left(\frac{1}{2}\nabla_{\bm{x}'}^2 - U(\bm{x}')\right)\right) + \frac{\gamma_1}{2}\left(\delta_{ij}\sigma^{ab}_y + i\sigma_{ij}^z \delta^{a\mathrm{q}}\delta^{b\mathrm{q}}\right)\right]\right.\nonumber\\
&+
\left.\sigma^y_{\sigma\sigma'} \gamma_{abc} \left[\chi^c(x')\,\sigma^{+}_{ij} + \bar{\phi}^c(x')\,\sigma^{-}_{ij}\right]\right\rbrace.
\end{align}

The equations for the field expectation values are obtained by varying the effective action with respect to the one-point functions. The fermionic expectation value, however, vanishes identically due to its Grassmann-odd character. At the same time, the Hubbard--Stratonovich field equations read
\begin{equation}
\frac{\delta S_{\mathrm{HS}}[\Psi,\Phi]}{\delta \Phi^{a}_{i\alpha}(x)} - \frac{i}{2} \Tr\left(\frac{\delta G_0^{-1}[\Phi]}{\delta \Phi^{a}_{i\alpha}(x)} G\right) = 0\,.
\end{equation}
Using Eqs.~\eqref{eq:S0} and  \eqref{eq:G0_inv}, one can readily obtain
\begin{equation}
\frac{\delta S_{\mathrm{HS}}}{\delta\chi^{\mathrm{q}}(x)} 
=
-\frac{\bar{\phi}^{\mathrm{cl}}(x)}{\bar{u}}\,,\qquad 
\frac{\delta S_{\mathrm{HS}}}{\delta\bar{\phi}^{\mathrm{q}}(x)} 
=
-\frac{\chi^{\mathrm{cl}}(x)}{u}\,,
\end{equation}
and
\begin{align}
i \Tr\left(\frac{\delta G_0^{-1}}{\delta \chi^{\mathrm{q}}(x)} G\right)
&=
\sigma^y_{\sigma\sigma'} \left[G^{\mathrm{clcl}}_{21;\sigma'\sigma}(x,x) + \frac{1}{4} G^{\mathrm{qq}}_{21;\sigma'\sigma}(x,x)\right]
=
2i \left\langle\psi^{\dagger}_{\uparrow}(x)\,\psi^{\dagger}_{\downarrow}(x) \right\rangle,\nonumber\\
i \Tr\left(\frac{\delta G_0^{-1}}{\delta \bar{\phi}^{\mathrm{q}}(x)} G\right)
&=
\sigma^y_{\sigma\sigma'} \left[G^{\mathrm{clcl}}_{12;\sigma'\sigma}(x,x) + \frac{1}{4} G^{\mathrm{qq}}_{12;\sigma'\sigma}(x,x)\right]
=
2i \left\langle\vphantom{\psi^{\dagger}_{\downarrow}}\psi_{\uparrow}(x)\,\psi_{\downarrow}(x) \right\rangle,
\end{align}
where we have used that, for physical configurations, $O^{\mathrm{q}} = M^{\mathrm{qq}} = 0$. Together, this yields
\begin{equation}
\chi^{\mathrm{cl}}(x) = -iu \left\langle\hat{\psi}_{\uparrow}(x)\,\hat{\psi}_{\downarrow}(x) \right\rangle,\qquad
\bar{\phi}^{\mathrm{cl}}(x) = -i\bar{u} \left\langle\hat{\psi}^{\dagger}_{\uparrow}(x)\,\hat{\psi}^{\dagger}_{\downarrow}(x) \right\rangle.
\end{equation}
reflecting the composite nature of the Hubbard--Stratonovich fields. In this work, we assume the absence of superconducting order, so that all anomalous (Nambu off-diagonal) components of the Hubbard--Stratonovich correlation functions vanish: $\langle\hat{\chi}^{(\dagger)}\rangle = \langle\hat{\phi}^{(\dagger)}\rangle = \langle\hat{\phi} \hat{\chi} \rangle = \ldots = 0$.

The propagator equations can be similarly obtained by varying the effective action with respect to $G$ and $D$:
\begin{equation}
\label{eq:Dyson}
\frac{\delta\Gamma}{\delta G} = 0:\; G^{-1} = G_0^{-1} - \Pi[G,D]\,,\qquad
\frac{\delta\Gamma}{\delta D} = 0:\; D^{-1} = D_0^{-1} - \Sigma[G,D]\,,
\end{equation}
where $\Pi$ and $\Sigma$ are the proper self-energies, defined as
\begin{equation}
\Pi^{ab}_{ij;\sigma\sigma'}(x,x') = -2i\frac{\delta \Gamma_2[G,D]}{\delta G^{ba}_{ji;\sigma'\sigma}(x',x)}\,,\qquad 
\Sigma^{ab}_{ij;\alpha\beta}(x,x') = 2i\frac{\delta \Gamma_2[G,D]}{\delta D^{ba}_{ji;\beta\alpha}(x',x)}\,.
\end{equation}
For what follows, it is convenient to rewrite the Dyson equations \eqref{eq:Dyson} in the equivalent form
\begin{equation}
\label{eq:Dyson_matrix}
\left[\left(G_0^{-1} - \Pi\right)\circ G\right]^{ab}_{ij;\sigma\sigma'}(x,x') = \delta^{ab}\delta_{ij}\delta_{\sigma\sigma'}\delta(x-x')\,,\qquad
\left[\left(D_0^{-1} - \Sigma\right)\circ D\right]^{ab}_{ij;\alpha\beta}(x,x') = \delta^{ab}\delta_{ij}\delta_{\alpha\beta}\delta(x-x')\,,
\end{equation}
or the adjoint version
\begin{equation}
\label{eq:Dyson_matrix_adjoint}
\left[G \circ \left(G_0^{-1} - \Pi\right)\right]^{ab}_{ij;\sigma\sigma'}(x,x') = \delta^{ab}\delta_{ij}\delta_{\sigma\sigma'}\delta(x-x')\,,\qquad
\left[D \circ \left(D_0^{-1} - \Sigma\right)\right]^{ab}_{ij;\alpha\beta}(x,x') = \delta^{ab}\delta_{ij}\delta_{\alpha\beta}\delta(x-x')\,,
\end{equation}
where $\circ$ denotes integration and summation over all variables and indices.

To close the Dyson equations, it remains to specify the self-energies $\Pi$ and $\Sigma$, which amounts to specifying $\Gamma_2$. Truncated at leading order in the loop expansion, it reads
\begin{equation}
\Gamma_2
=
\frac{i}{2}\gamma_{abc}\gamma_{def}\int_{x,x'} D_{\bar{i}j;ij}^{cf}(x,x')\left[G^{ad}_{i\bar{j};\uparrow\uparrow}(x,x')\,G^{be}_{i\bar{j};\downarrow\downarrow}(x,x') - G^{ae}_{i\bar{j};\uparrow\downarrow}(x,x')\,G^{bd}_{i\bar{j};\downarrow\uparrow}(x,x') \right],
\end{equation}
 where we have adopted the short-hand notation $\bar{j} \equiv 3 -j $, $x \equiv (x^0,\bm{x}) = (t,\bm{x})$, and $\int_x \equiv \int\dd{t}\dd[d]{\bm{x}}$.

In this work, we assume the initial condition is chosen invariant under global $\mathrm{SU}(2)$ spin rotations, and such correlations are not dynamically generated as the density-density interaction conserves the total spin. This assumes that the spin-symmetric state is stable against spin-triplet fluctuations, which in equilibrium amounts to the interaction lying below the Stoner and spin-triplet Pomeranchuk thresholds \cite{BaymPethick1991}. As a result, the fermionic one-body density matrix $\langle\mathcal{T}_{\mathcal{C}}\,\psi^\dagger_\sigma(x)\psi_{\sigma'}(x')\rangle$ is assumed to be spin-diagonal and unpolarized, $\langle\mathcal{T}_{\mathcal{C}}\,\psi^\dagger_\sigma(x)\psi_{\sigma'}(x')\rangle \propto \delta_{\sigma\sigma'} $. Consequently, the relevant nonvanishing self-energy entries read
\begin{align}
\label{eq:self_energies_clq}
\Pi^{\mathrm{qq}}_{11;\sigma\sigma}(x,x')
&=
-\left[D^{\mathrm{clcl}}_{11;21}(x,x')\,G^{\mathrm{clcl}}_{22;\bar{\sigma}\bar{\sigma}}(x,x') + \frac{1}{4}\left(D^{\mathrm{clq}}_{11;21}(x,x')\,G^{\mathrm{clq}}_{22;\bar{\sigma}\bar{\sigma}}(x,x') + D^{\mathrm{qcl}}_{11;21}(x,x')\,G^{\mathrm{qcl}}_{22;\bar{\sigma}\bar{\sigma}}(x,x')\right)\right],\nonumber\\
\Pi^{\mathrm{qcl}}_{11;\sigma\sigma}(x,x')
&=
-\left[D^{\mathrm{clq}}_{11;21}(x,x')\,G^{\mathrm{clcl}}_{22;\bar{\sigma}\bar{\sigma}}(x,x') + D^{\mathrm{clcl}}_{11;21}(x,x')\,G^{\mathrm{clq}}_{22;\bar{\sigma}\bar{\sigma}}(x,x')\right],\nonumber\\
\Pi^{\mathrm{clq}}_{11;\sigma\sigma}(x,x')
&=
-\left[D^{\mathrm{qcl}}_{11;21}(x,x')\,G^{\mathrm{clcl}}_{22;\bar{\sigma}\bar{\sigma}}(x,x') + D^{\mathrm{clcl}}_{11;21}(x,x')\,G^{\mathrm{qcl}}_{22;\bar{\sigma}\bar{\sigma}}(x,x')\right],\\
\Sigma^{\mathrm{qq}}_{11;12}(x,x')
&=
-\left[G^{\mathrm{clcl}}_{11;\uparrow\uparrow}(x,x')\,G^{\mathrm{clcl}}_{11;\downarrow\downarrow}(x,x') + \frac{1}{4}\left(G^{\mathrm{clq}}_{11;\uparrow\uparrow}(x,x')\,G^{\mathrm{clq}}_{11;\downarrow\downarrow}(x,x') + G^{\mathrm{qcl}}_{11;\uparrow\uparrow}(x,x')\,G^{\mathrm{qcl}}_{11;\downarrow\downarrow}(x,x')\right)\right],\nonumber\\
\Sigma^{\mathrm{qcl}}_{11;12}(x,x')
&=
-\left[G^{\mathrm{clq}}_{11;\uparrow\uparrow}(x,x')\,G^{\mathrm{clcl}}_{11;\downarrow\downarrow}(x,x') + G^{\mathrm{clcl}}_{11;\uparrow\uparrow}(x,x')\,G^{\mathrm{clq}}_{11;\downarrow\downarrow}(x,x')\right],\nonumber\\
\Sigma^{\mathrm{clq}}_{11;12}(x,x')
&=
-\left[G^{\mathrm{qcl}}_{11;\uparrow\uparrow}(x,x')\,G^{\mathrm{clcl}}_{11;\downarrow\downarrow}(x,x') + G^{\mathrm{clcl}}_{11;\uparrow\uparrow}(x,x')\,G^{\mathrm{qcl}}_{11;\downarrow\downarrow}(x,x')\right].
\end{align}
The remaining components can be obtained using the symmetry relations $\Pi^{ab}_{22;\sigma\sigma}(x,x') = -\Pi^{ba}_{11;\sigma\sigma}(x',x)$ and $\Sigma^{ab}_{22;21}(x,x') = \Sigma^{ba}_{11;12}(x',x)$. In the above expressions, we have omitted terms involving $G^{\mathrm{qq}}$, which vanishes for $t\neq t'$. At coinciding times, however, the identity $G^{++}+G^{--}=G^{+-}+G^{-+}$ underlying $G^{\mathrm{qq}}=0$ breaks down, since the $\bar{\psi}$ fields of the time-sliced action are evaluated one timestep later along the contour~\cite{Kamenev2011,Sieberer:2023qyv}. This affects only the time-local parts of the self-energies, which we discuss in detail below.

With these symmetry considerations, the only nontrivial propagators are $G_{11;\sigma\sigma}$ and $D_{11;21}$, which we denote as $G_{\sigma}$ and $D$, respectively. Likewise, we write $\Pi_{\sigma}\equiv\Pi_{11;\sigma\sigma}$ and $\Sigma\equiv\Sigma_{11;12}$ for the corresponding self-energies, while all remaining components follow from the symmetry relations. We further note that the HS propagators contain a singular, spacetime-local part, which coincides with the bare propagator $D_0$. This reflects the auxiliary, vertex-like nature of the HS fields, which carry no dynamics of their own. It is then suggestive to remove this part and to do the same for the fermionic self-energies,
\begin{align}
D^{\mathrm{clcl}}(x,x') &= D^F(x,x') + \gamma_2\delta(x-x')\,, &\Pi^{\mathrm{qq}}_{\sigma}(x,x')  &= -\mathcal{K}_{\sigma}(x)\,\delta(x-x') + \Pi^F_{\sigma}(x,x')\,,\nonumber\\
D^{\mathrm{clq}}(x,x') &= -i D^R(x,x')  - iu\,\delta(x-x')\,, &\Pi^{\mathrm{clq}}_{\sigma}(x,x') &= -i\left[\Pi_{\sigma}^{(0)}(x) + i\mathcal{K}_{\sigma}(x)\right]\delta(x-x') - i\Pi^A_{\sigma}(x,x')\,,\nonumber\\
D^{\mathrm{qcl}}(x,x') &= -i D^A(x,x') - i\bar{u}\,\delta(x-x')\,, &\Pi^{\mathrm{qcl}}_{\sigma}(x,x') &= -i\left[\Pi_{\sigma}^{(0)}(x) - i\mathcal{K}_{\sigma}(x)\right]\delta(x-x') - i\Pi^R_{\sigma}(x,x')\,,
\end{align}
where $\Pi^{(0)}_{\sigma}$ and $\mathcal{K}_{\sigma}$ act as a local energy shift and a local loss rate, respectively. In contrast, $G_{\sigma}$ and $\Sigma$ contain no such spacetime-singular contributions, and we simply parameterize them as
\begin{align}
G_{\sigma}^{\mathrm{clcl}}(x,x') &= G_{\sigma}^F(x,x')\,,&
G_{\sigma}^{\mathrm{clq}}(x,x') &= -i G^R_{\sigma}(x,x')\,,&
G_{\sigma}^{\mathrm{qcl}}(x,x') &= -i G^A_{\sigma}(x,x')\,,\nonumber\\
\Sigma^{\mathrm{qq}}(x,x')  &= \Sigma^F(x,x')\,,&
\Sigma^{\mathrm{clq}}(x,x') &= -i\Sigma^A(x,x')\,,&
\Sigma^{\mathrm{qcl}}(x,x') &= -i\Sigma^R(x,x')\,.
\end{align}
Using this parametrization, the Dyson equations \eqref{eq:Dyson_matrix} can be explicitly written as
\begin{align}
\label{eq:KB_fermions}
\left[i\partial_t + \frac{1}{2}\nabla_{\bm{x}}^2 - \mathcal{U}_{\sigma}(x) + i \Gamma_{\sigma}(x)\right] G^F_{\sigma}(x,x') &= -\Gamma_{\sigma}(x)\,G^A_{\sigma}(x,x')  + \int_{x''} \left[\Pi^R_{\sigma}(x,x'')\,G^F_{\sigma}(x'',x') + \Pi^F_{\sigma}(x,x'')\,G^A_{\sigma}(x'',x')\right],\nonumber\\
\left[i\partial_t + \frac{1}{2}\nabla_{\bm{x}}^2 - \mathcal{U}_{\sigma}(x) + i \Gamma_{\sigma}(x)\right] G^R_{\sigma}(x,x') &= -\delta(x-x') + \int_{x''} \Pi^R_{\sigma}(x,x'')\,G^R_{\sigma}(x'',x')\,,\nonumber\\
\left[i\partial_t + \frac{1}{2}\nabla_{\bm{x}}^2 - \mathcal{U}_{\sigma}(x) - i \Gamma_{\sigma}(x)\right] G^A_{\sigma}(x,x') &= -\delta(x-x') + \int_{x''} \Pi^A_{\sigma}(x,x'')\,G^A_{\sigma}(x'',x')\,,
\end{align}
and
\begin{align}
\label{eq:KB_HS}
D^F(x,x') 
&=
\frac{\gamma_2}{\bar{u}} D^A(x,x') - |u|^2 \Sigma^F(x,x') - \gamma_2u \Sigma^R(x,x')
-
u\int_{x''}\left[\Sigma^R(x,x'')\,D^F(x'',x') + \Sigma^F(x,x'')\,D^A(x'',x')\right],\nonumber\\
D^R(x,x') 
&=
-u^2 \Sigma^R(x,x') - u\int_{x''}\Sigma^R(x,x'')\,D^R(x'',x')\,,\nonumber\\
D^A(x,x') 
&=
-\bar{u}^2\Sigma^A(x,x') - \bar{u}\int_{x''}\Sigma^A(x,x'')\,D^A(x'',x')\,.
\end{align}
Here, we have introduced the notation $\Gamma_{\sigma}(x) \equiv \gamma_1/2 + \mathcal{K}_{\sigma}(x)$ and $\mathcal{U}_{\sigma}(x) \equiv U(\bm{x}) + \Pi^{(0)}_{\sigma}(x)$. Note that Eq.~\eqref{eq:KB_fermions} holds component-wise, with no summation over repeated spin indices. 

Written in this time-differential form, the Dyson equations \eqref{eq:KB_fermions} are known as the Kadanoff--Baym equations \cite{Danielewicz:1982kk,Rammer:2007,Calzetta:2008iqa} and form the basis of the kinetic theory derived in the next section. In turn, the Dyson equations \eqref{eq:KB_HS} determine the Hubbard--Stratonovich propagators, which enter the fermionic self-energies and, as will be shown below, play the role of the self-consistent $T$-matrix.

\subsection{Kinetic theory}
The Kadanoff--Baym equations discussed in the previous section capture the full nonequilibrium dynamics within the chosen truncation, including memory effects and off-shell processes. Once the system has evolved beyond the initial transient, dephasing and interaction-induced damping suppress long-range correlations in the relative coordinates. In this regime, the correlation functions then typically vary only weakly with the central coordinates $X=(x+x')/2$ compared with their dependence on the relative coordinates $s=x-x'$. This separation of scales allows for a controlled expansion in gradients $\partial_X$, known as the gradient expansion \cite{Knoll:2001jx,Rammer:1986zz,Berges:2015kfa}. This expansion is conveniently organized in Wigner space, obtained by Fourier transforming correlation functions with respect to the fast coordinate $s$ while retaining the slowly varying central coordinate $X$:
\begin{equation}
O_W(X,p) = \int_{-2 X^0}^{2 X^0} \dd{s^0} \int \dd[d]{\bm{s}} \mathrm{e}^{i p \cdot s}\,O\left(X + s/2, X - s/2\right),
\end{equation}
where $p \cdot s = p_{\mu} s^{\mu} \equiv  p^0 s^0 - \bm{p}\cdot\bm{s}$. In the following, we drop the subscript $W$ denoting the Wigner transform whenever there is no risk of confusion.

To proceed, it is convenient to supplement Eqs.~\eqref{eq:KB_fermions} with the adjoint Kadanoff--Baym equations, which follow from Eq.~\eqref{eq:Dyson_matrix_adjoint}:
\begin{align}
\label{eq:KB_fermions_adjoint}
\left[-i\partial_{t'} + \frac{1}{2}\nabla_{\bm{x}'}^2 - \mathcal{U}_{\sigma}(x') - i \Gamma_{\sigma}(x')\right] G^F_{\sigma}(x,x') &= -\Gamma_{\sigma}(x')\,G^R_{\sigma}(x,x')  + \int_{x''} \left[G^F_{\sigma}(x,x'')\,\Pi^A_{\sigma}(x'',x') + G^R_{\sigma}(x,x'')\,\Pi^F_{\sigma}(x'',x')\right],\nonumber\\
\left[-i\partial_{t'} + \frac{1}{2}\nabla_{\bm{x}'}^2 - \mathcal{U}_{\sigma}(x') + i \Gamma_{\sigma}(x')\right] G^R_{\sigma}(x,x') &= -\delta(x-x') + \int_{x''} G^R_{\sigma}(x,x'')\,\Pi^R_{\sigma}(x'',x')\,,\nonumber\\
\left[-i\partial_{t'} + \frac{1}{2}\nabla_{\bm{x}'}^2 - \mathcal{U}_{\sigma}(x') - i \Gamma_{\sigma}(x')\right] G^A_{\sigma}(x,x') &= -\delta(x-x') + \int_{x''} G^A_{\sigma}(x,x'')\,\Pi^A_{\sigma}(x'',x')\,.
\end{align}
Subtracting Eqs.~\eqref{eq:KB_fermions_adjoint} from Eqs.~\eqref{eq:KB_fermions} and using that $(A\circ B)_W = A_W B_W + O(\partial_X\cdot\partial_p)$ \cite{Rammer:1986zz,Kamenev2011}, we obtain, to leading order in the gradient expansion,
\begin{align}
\label{eq:grad_exp_1}
&\left(\partial_t + \bm{p}\cdot\nabla_{\bm{r}} + \partial_X\mathcal{U}_{\sigma}\cdot\partial_p\right) G^F_{\sigma}(X,p) + 2\Gamma_{\sigma}(X) \left[G_{\sigma}^F(X,p) + i \rho_{\sigma}(X,p)/2\right] = C_{\sigma}^{\mathrm{coll}}(X,p)\,,\nonumber\\
&\left(\partial_t + \bm{p}\cdot\nabla_{\bm{r}} + \partial_X\mathcal{U}_{\sigma}\cdot\partial_p\right) \rho_{\sigma}(X,p) = 0\,.
\end{align}
Here, $t \equiv X^0$, $\bm{r} \equiv \bm{X}$, $\rho_{\sigma} = G^R_{\sigma} - G^A_{\sigma}$, and
\begin{equation}
C_{\sigma}^{\mathrm{coll}}(X,p) = -i\left[\Pi^{\rho}_{\sigma}(X,p)\,G^F_{\sigma}(X,p) - \Pi^F_{\sigma}(X,p)\,\rho_{\sigma}(X,p)\right],
\end{equation}
with $\Pi^{\rho}_{\sigma} = \Pi^R_{\sigma} - \Pi^A_{\sigma}$. Hence, to leading order in the gradient expansion, the spectral function evolves collisionlessly, driven only by local terms. In contrast, the self-energies survive in the equation for the statistical function $G^F_{\sigma}$ already at this order, giving rise to a collision integral, which, together with the local loss rate $\Gamma_\sigma$, governs the redistribution and decay of the particles. 

The combination of $G^F$ and $\rho$ appearing in Eq.~\eqref{eq:grad_exp_1} suggests a useful parameterization of the statistical propagator:
\begin{equation}
\label{eq:KBA}
G^F_{\sigma}(X,p) = i\left[\,\mathfrak{f}_{\sigma}(X,p) - 1/2\right]\rho_{\sigma}(X,p)\,.
\end{equation}
Written in this form this parameterization is exact, as it merely defines $\mathfrak{f}_{\sigma}(X,p)$. In thermal equilibrium, it takes the form of the fluctuation--dissipation relation, with $n_{\sigma}$ given by the Fermi--Dirac distribution and weighted by the spectral function $\rho_{\sigma}$ around the mass shell. This motivates defining the nonequilibrium phase-space distribution as the spectral average
\begin{equation} 
f_{\sigma}(t,\bm{r},\bm{p}) \equiv \int_{-\infty}^{\infty}\frac{\dd{p^0}}{2\pi i}\,\rho_{\sigma}(X,p)\,\mathfrak{f}_{\sigma}(X,p)\,.
\end{equation}
In the literature, one also finds parameterizations in which the energy or the momentum dependence of $\mathfrak{f}_\sigma$ is neglected~\cite{Lipavsky:1986,Chantesana:2018qsb}, the former being commonly referred to as the Kadanoff--Baym ansatz. In the quasiparticle limit, only the on-shell value of $\mathfrak{f}_\sigma$ enters, such that the former becomes exact, while the latter requires $\mathfrak{f}_\sigma$ to depend on momentum only through the quasiparticle energy.

Multiplying the second equation of \eqref{eq:grad_exp_1} by $i/2$ and then adding it to the first equation yields
\begin{equation}
i\rho_{\sigma}(X,p) \left(\partial_t + \bm{p}\cdot\nabla_{\bm{r}} + \partial_X\mathcal{U}_{\sigma}\cdot\partial_p + 2 \Gamma_{\sigma}\right) \mathfrak{f}_{\sigma}(X,p) = C_{\sigma}^{\mathrm{coll}}(X,p)\,,
\end{equation}
where we have used the chain rule and the second equation in \eqref{eq:grad_exp_1}. Then, upon integrating it over $p^0$, one obtains
\begin{equation}
\left(\partial_t + \bm{p}\cdot\nabla_{\bm{r}} + \partial_X\mathcal{U}_{\sigma}\cdot\partial_p + 2 \Gamma_{\sigma}\right) f_{\sigma}(t,\bm{r},\bm{p}) = I_{\sigma}[\,\mathfrak{f}\,](t,\bm{r},\bm{p})\,,
\end{equation}
with 
\begin{equation}
I_{\sigma}[\,\mathfrak{f}\,](t,\bm{r},\bm{p}) \equiv -\int_{-\infty}^{\infty} \frac{\dd{p^0}}{2\pi}\,C_{\sigma}^{\mathrm{coll}}(X,p)
=
i\int_{-\infty}^{\infty} \frac{\dd{p^0}}{2\pi} \left[\Pi^{\rho}_{\sigma}(X,p)\,G^F_{\sigma}(X,p) - \Pi^F_{\sigma}(X,p)\,\rho_{\sigma}(X,p)\right],
\end{equation}
To make the gain--loss structure of the collision integral explicit, we rewrite its integrand as
\begin{align}
i\left(\Pi_{\sigma}^{\rho}\,G^F_{\sigma} - \Pi_{\sigma}^F \rho_{\sigma}\right)
&=
\left(\Pi_{\sigma}^F + \frac{i}{2}\Pi_{\sigma}^{\rho}\right) \left(G^F_{\sigma} - \frac{i}{2}\rho_{\sigma}\right) - \left(\Pi^F_{\sigma} - \frac{i}{2}\Pi_{\sigma}^{\rho}\right)\left(G^F_{\sigma} + \frac{i}{2}\rho_{\sigma}\right)\nonumber\\
&=
i \rho_{\sigma} \left[\left(\Pi_{\sigma}^F + \frac{i}{2}\Pi_{\sigma}^{\rho}\right)\left(\vphantom{\frac{i}{2}}\mathfrak{f}_{\sigma} - 1\right) - \left(\Pi^F_{\sigma} - \frac{i}{2}\Pi_{\sigma}^{\rho}\right) \mathfrak{f}_{\sigma}\right].
\end{align}
Turning to the self-energies, we note that they are pointwise products of two-point functions in coordinate space, so that their Wigner transforms are convolutions in momentum space \cite{Kamenev2011}. Using $G^{ab}_{22;\sigma\sigma}(x,x')=-G^{ba}_{\sigma}(x',x)$, we obtain
\begin{align}
\Pi^F_{\sigma}(X,p)
&=
\int_q \left[D^F(X,q)\,G^F_{\bar{\sigma}}(X,q-p) + \frac{1}{4}D^{\rho}(X,q)\,\rho_{\bar{\sigma}}(X,q-p)\right],\nonumber\\
\Pi^{\rho}_\sigma(X,p)
&=
\int_q\left[D^{\rho}(X,q)\,G^F_{\bar{\sigma}}(X,q-p)-D^F(X,q)\,\rho_{\bar{\sigma}}(X,q-p)\right],\\
\Sigma^F(X,p)
&=
-\int_q\left[G^F_{\uparrow}(X,q)\,G^F_{\downarrow}(X,p-q)-\frac{1}{4} \rho_{\uparrow}(X,q)\,\rho_{\downarrow}(X,p-q)\right],\nonumber\\
\Sigma^{\rho}(X,p)
&=
-\int_q\left[G^F_{\uparrow}(X,q)\,\rho_{\downarrow}(X,p-q)+\rho_{\uparrow}(X,q)\,G^F_\downarrow(X,p-q)\right].
\label{eq:sm_self_energies}
\end{align}
Note that the retarded and advanced components $\Sigma^{R(A)}$ follow from the expression of $\Sigma^{\rho}$ by replacing $\rho \to G^{R(A)}$. Taking into account the coinciding-time contributions discussed after Eq.~\eqref{eq:self_energies_clq}, the time-local contributions read
\begin{align}
\Pi^{(0)}_{\sigma}(X)
&=
g \left[G^{\mathrm{clcl}}_{\bar{\sigma}}(x,x) + \frac{1}{4} G^{\mathrm{qq}}_{\bar{\sigma}}(x,x)\right] + \frac{i\gamma_2}{2}\left[G^{\mathrm{clq}}_{\bar{\sigma}}(x,x) + G^{\mathrm{qcl}}_{\bar{\sigma}}(x,x)\right]\nonumber\\
&=
\frac{g}{2}\left[G^{++}_{\bar{\sigma}}(x,x) + G^{--}_{\bar{\sigma}}(x,x)\right] + \frac{i\gamma_2}{2}\left[G^{++}_{\bar{\sigma}}(x,x) - G^{--}_{\bar{\sigma}}(x,x)\right]
= g\,G^{+-}_{\bar{\sigma}}(x,x) = -g n_{\bar{\sigma}}(X)\,,\nonumber\\
\mathcal{K}_{\sigma}(X) 
&=
-\gamma_2 \left[G^{\mathrm{clcl}}_{\bar{\sigma}}(x,x) - \frac{1}{4} G^{\mathrm{qq}}_{\bar{\sigma}}(x,x) + \frac{1}{2} \left(G^{\mathrm{qcl}}_{\bar{\sigma}}(x,x) - G^{\mathrm{clq}}_{\bar{\sigma}}(x,x)\right)\right]\nonumber\\
&=
-\frac{\gamma_2}{2}\left[G^{+-}_{\bar{\sigma}}(x,x) + G^{-+}_{\bar{\sigma}}(x,x) - G^{+-}_{\bar{\sigma}}(x,x) + G^{-+}_{\bar{\sigma}}(x,x)\right] =
-\gamma_2\, G^{+-}_{\bar{\sigma}}(x,x) = \gamma_2 n_{\bar{\sigma}}(X)\,,
\end{align}
where $n_{\sigma}(t,\bm{r}) = \int_{\bm{p}}f(t,\bm{r},\bm{p})$ is the time-dependent local density of fermions with spin $\sigma$. Here, we have used that, at coinciding times, $G^{++}_{\bar{\sigma}}(x,x)=G^{--}_{\bar{\sigma}}(x,x)=G^{+-}_{\bar{\sigma}}(x,x)=-\langle\hat{\psi}^{\dagger}_{\bar{\sigma}}(x)\hat{\psi}_{\bar{\sigma}}(x)\rangle$, since the $\bar{\psi}$ fields sit one timestep later on the contour. 

The combinations appearing in the collision integral now read
\begin{equation}
\Pi^F_{\sigma}(X,p) \pm \frac{i}{2}\Pi^{\rho}_{\sigma}(X,p) 
=
\int_q\left[D^F(X,q+p) \pm \frac{i}{2}D^{\rho}(X,q+p)\right]\left[G^F_{\bar{\sigma}}(X,q) \mp \frac{i}{2}\rho_{\bar{\sigma}}(X,q)\right].
\end{equation}
To leading order in the gradient expansion, the Dyson equations for the Hubbard--Stratonovich propagators become algebraic, which allows us to solve them in closed form,
\begin{equation}
D^R = T^R - u\,,\qquad D^A = T^A - \bar{u}\,,\qquad D^F = -T^R\,\Sigma^F\,T^A - \gamma_2 \left[1 - \frac{T^R\,T^A}{|u|^2}\right],
\end{equation}
where all the functions are evaluated at $(X,p)$ and we have introduced
\begin{equation}
T^R(X,p) \equiv \frac{u}{1 + u\,\Sigma^R(X,p)}\,,\qquad 
T^A(X,p) \equiv \frac{\bar{u}}{1 + \bar{u}\,\Sigma^A(X,p)} = T^R(X,p)^{*}\,.
\label{eq:T_matrices}
\end{equation}
With this, we obtain
\begin{equation}
D^F + \frac{i}{2} D^{\rho} = -T^R \left(\Sigma^F+\frac{i}{2}\Sigma^\rho\right) T^A,\qquad
D^F - \frac{i}{2}D^{\rho} =-T^R\left(\Sigma^F-\frac{i}{2}\Sigma^\rho\right)T^A - 2\gamma_2 \left(1-\frac{T^R T^A}{|u|^2}\right),
\end{equation}
where, again, all the functions are evaluated at $(X,p)$. Using the parameterization \eqref{eq:KBA} we can write
\begin{align}
\Sigma^F(X,p) + \frac{i}{2}\Sigma^{\rho}(X,p)
&=
\int_q \mathfrak{f}_{\uparrow}(X,q)\,\mathfrak{f}_{\downarrow}(X,p-q)\,\rho_{\uparrow}(X,q)\,\rho_{\downarrow}(X,p-q)\,,\nonumber\\
\Sigma^F(X,p) - \frac{i}{2}\Sigma^{\rho}(X,p)
&=
\int_q \left[1 - \mathfrak{f}_{\uparrow}(X,q)\right]\left[1 - \mathfrak{f}_{\downarrow}(X,p-q)\right]\rho_{\uparrow}(X,q)\,\rho_{\downarrow}(X,p-q)\,.
\end{align}
Assembling the pieces together, we get
\begin{align}
\label{eq:PiF_rho_comb}
\Pi^F_{\sigma}(X,p) + \frac{i}{2}\Pi^\rho_{\sigma}(X,p)
&=
i\int_{q,p',q'} \left|T^R(X,p+q)\right|^2 (2\pi)^{d+1}\delta^{(d+1)}(p+q-p'-q')  \rho_{\sigma p'} \rho_{\bar{\sigma}q'} \rho_{\bar{\sigma}q}\, \mathfrak{f}_{\sigma p'} \mathfrak{f}_{\bar{\sigma}q'}\left(1 - \mathfrak{f}_{\bar{\sigma}q}\right)\,,\nonumber\\
\Pi^F_{\sigma}(X,p) - \frac{i}{2}\Pi^\rho_{\sigma}(X,p)
&=
-i\int_{q,p',q'} \left|T^R(X,p+q)\right|^2 (2\pi)^{d+1}\delta^{(d+1)}(p+q-p'-q')  \rho_{\sigma p'} \rho_{\bar{\sigma}q'} \rho_{\bar{\sigma}q} \left(1 - \mathfrak{f}_{\sigma p'}\right)\left(1 - \mathfrak{f}_{\bar{\sigma}q'}\right) \mathfrak{f}_{\bar{\sigma}q}\nonumber\\
&-
2i\gamma_2\int_l \left[1 - \frac{\left|T^R(X,p+q)\right|^2}{|u|^2}\right] \rho_{\bar{\sigma} q} \mathfrak{f}_{\bar{\sigma} q}\,,
\end{align}
where we have introduced the short-hand notation $\mathfrak{f}_{\sigma p} \equiv \mathfrak{f}_{\sigma}(X,p)$ and $\rho_{\sigma p} \equiv \rho_{\sigma}(X,p)$. Finally, plugging this into the collision integral yields
\begin{equation}
\label{eq:boltzmann}
\left(\partial_t + \bm{p}\cdot\nabla_{\bm{r}} + \partial_X\mathcal{U}_{\sigma}\cdot\partial_p + \gamma_1 \right) f_{\sigma\bm{p}} = I^{\mathrm{in}}_{\sigma}[\,\mathfrak{f}\,] + I^{\mathrm{el}}[\,\mathfrak{f}\,]\,,
\end{equation}
where the collision kernels take the form
\begin{align}
\label{eq:I_off_shell}
I_{\sigma}^{\mathrm{in}}[f] 
&=
\frac{2 \gamma_2}{|u|^2}\int\frac{\dd{p^0}}{2\pi}\int_q \left|T^R(X,p+q)\right|^2 \rho_{\sigma p}\rho_{\bar{\sigma}q}\,\mathfrak{f}_{\sigma p}\mathfrak{f}_{\bar{\sigma}q}\,,\nonumber\\
I_{\sigma}^{\mathrm{el}}[f] 
&=
\int\dd{\Omega_{\sigma}(p,q,p',q')}\left|T^R(X,p+q)\right|^2 \left[\left(1 - \mathfrak{f}_{\sigma p}\right)\left(1 - \mathfrak{f}_{\bar{\sigma}q}\right) \mathfrak{f}_{\sigma p'} \mathfrak{f}_{\bar{\sigma} q'} - \mathfrak{f}_{\sigma p}\mathfrak{f}_{\bar{\sigma}q} \left(1 - \mathfrak{f}_{\sigma p'}\right)\left(1 - \mathfrak{f}_{\bar{\sigma} q'}\right)\right],
\end{align}
where we have introduced the notation
\begin{equation}
\int \dd{\Omega_{\sigma}(p,q,p',q')}
\equiv
\int\frac{\dd{p^0}\dd{q^0}\dd{p'^0}\dd{q'^0}}{(2\pi)^{4-(d+1)}}\int_{\bm{q},\bm{p},\bm{q'}}\delta^{(d+1)}(p+q-p'-q')\,\rho_{\sigma p} \rho_{\bar{\sigma}q} \rho_{\sigma p'} \rho_{\bar{\sigma} q'}\,.
\end{equation}
Note that the two-body part of the dissipative tadpole contribution, $\mathcal{K}_{\sigma}$, is cancelled by the first term in the square brackets in the last line of Eq.~\eqref{eq:PiF_rho_comb}, such that the pair loss enters only through the in-medium $T$-matrix.  

Equation \eqref{eq:boltzmann} with collision kernels \eqref{eq:I_off_shell} constitutes an off-shell generalization of the Boltzmann equation: it does not require the excitations to lie precisely on the mass shell and accounts for the finite lifetime of quasiparticles. In dissipative gases such as the one considered in this work, this lifetime is finite even in the absence of interactions, being set by the inverse loss rates. However, provided these losses are weak and the interaction-induced broadening is small compared to the energy scales on which the distributions vary, the spectral functions are sharply peaked, and one may resort to the quasiparticle approximation,
\begin{equation} 
\rho_{\sigma}(X,p) \simeq 2\pi i \,\delta\left(p^0 - \epsilon_{\sigma}(X,\bm{p})\right),\qquad \epsilon_{\sigma}(X,\bm{p}) = \frac{\bm{p}^2}{2} + U(\bm{r}) - gn_{\bar{\sigma}}(t,\bm{r})\,. 
\end{equation}
In this limit, the spectral functions merely put frequencies on the mass shell, collapsing the frequency integrals in Eqs.~\eqref{eq:I_off_shell}. The result is the conventional Boltzmann equation, with the on-shell collision kernels:
\begin{align}
\label{eq:I_on_shell}
I_{\sigma}^{\mathrm{in}}[f] 
&=
-\frac{2\gamma_2}{|u|^2}\int_{\bm{q}} \left|T^R(X;\epsilon_{\sigma\bm{p}} + \epsilon_{\bar{\sigma}\bm{q}},\bm{p}+\bm{q})\right|^2\,f_{\sigma \bm{p}}f_{\bar{\sigma}\bm{q}}\,,\nonumber\\
I_{\sigma}^{\mathrm{el}}[f] 
&=
\int \dd{\Omega_{\sigma}(p,q,p',q')}  \left|T^R(X;\epsilon_{\sigma\bm{p}} + \epsilon_{\bar{\sigma}\bm{q}},\bm{p}+\bm{q})\right|^2
\left[\left(1 - f_{\sigma \bm{p}}\right)\left(1 - f_{\bar{\sigma}\bm{q}}\right) f_{\sigma \bm{p'}} f_{\bar{\sigma} \bm{q'}} - f_{\sigma \bm{p}} f_{\bar{\sigma}\bm{q}} \left(1 - f_{\sigma \bm{p'}}\right)\left(1 - f_{\bar{\sigma} \bm{q'}}\right)\right],
\end{align}
where we have similarly introduced the short-hand notation $f_{\sigma\bm{p}} \equiv f_{\sigma}(t,\bm{r},\bm{p})$.

As a final remark, we note that the presented derivation carries over to both identical and two-component Bose gases with small modifications. For two-component Bose gases, the main modification is that all signs originating from the anticommutation of Grassmann fields disappear, and the Kadanoff--Baym parameterization reads $G^F=i(\,\mathfrak{f}+1/2)\rho$. Consequently, the Pauli-blocking factors $1-f$ in the collision integral and $1-f_{\uparrow}-f_{\downarrow}$ in the particle-particle bubble $\Sigma^R$ are replaced by the Bose-enhancement factors $1+f$ and $1+f_{\uparrow}+f_{\downarrow}$, respectively. In the nondegenerate regime considered in this work, $f_{\sigma\bm{p}}\ll 1$, these factors reduce to unity, such that the kinetic equation to leading order in the fugacity is identical for fermions and bosons, cf. also the discussion in the next section. For single component Bose gases, one needs to take into account the in-distinguishability of the bosons in the loops which leads to trivial factors of $2$. In any case, the structure of the $T$-matrix and of the loss term remains unchanged. 

\subsection{Dilute limit and vacuum renormalization}
In the high-temperature limit, the gas is dilute with a small phase space density \mbox{$n_{\sigma}\lambda_T^3\ll1$}, where \mbox{$n_{\sigma}= \int_{\bm p} f_{\sigma, {\bm p}}$} is the density of particles with spin $\sigma$ and \mbox{$\lambda_T$} is the thermal de Broglie wavelength. Thus, at high temperatures, we can perform a virial expansion of the collision integral and retain only terms to leading order in the fugacity \mbox{$z =  n_{\sigma}\lambda_T^3 \ll 1$}. In this limit, the distribution functions are proportional to \mbox{$z$} at leading order.

We first consider the retarded two-body in  Eq.~\eqref{eq:T_matrices}. To leading order in the virial expansion, we can neglect the many-body background in evaluating the retarded self-energy. This is tantamount to neglecting how Pauli blocking (or Bose enhancement, in the case of bosons) modify the scattering of pairs of particles, which is valid in the dilute limit to leading order. In this case, the retarded self-energy in Eq.~\eqref{eq:sm_self_energies} reads 
\begin{equation}
\Sigma^{R}(X,p) 
=
\frac{i}{2}\int_q\left[\rho_{\uparrow}(X,q)\,G^R_{\downarrow}(X,p-q) + G^R_{\uparrow}(X,q)\,\rho_{\downarrow}(X,p-q)\right]\nonumber\\
=
i\int_q G^R_{\uparrow}(X,q)\,G^R_{\downarrow}(X,p-q)\,.
\end{equation}
Using the bare expression for $G^R$, one readily obtains
\begin{equation}
\Sigma^{R}(X,p)
=
i\int_q \frac{1}{\left(q^0 - \omega_{\bm{q}} + i\gamma_1/2\right)\left(p^0 - q^0 - \omega_{\bm{p}-\bm{q}} + i\gamma_1/2\right)}
=
\int\frac{\dd[d]{\bm{q}}}{(2\pi)^d} \frac{1}{p^0 - \omega_{\bm{q}} - \omega_{\bm{p}-\bm{q}} + i\gamma_1}\,,
\end{equation}
where $\omega_{\bm{p}} \equiv \bm{p}^2/2 + U(\bm{r})$ is the bare dispersion relation. The $T$-matrix \eqref{eq:T_matrices} in the dilute limit therefore reads
\begin{equation}
\left[T^R(X,p)\right]^{-1} = \frac{1}{u} + \int\frac{\dd[d]{\bm{q}}}{(2\pi)^d} \frac{1}{p^0 - \omega_{\bm{q}} - \omega_{\bm{p}-\bm{q}} + i\gamma_1},
\label{eq:vacuum_T_matrix_calc}
\end{equation}
In $d=3$ spatial dimensions, the integral in Eq.~\eqref{eq:vacuum_T_matrix_calc} is formally divergent and needs regularization. To that end we introduce an ultraviolet cutoff scale $\Lambda$ and evaluate Eq.~\eqref{eq:vacuum_T_matrix_calc} explicitly
\begin{equation}
\left[T^R(X,p)\right]^{-1} = \frac{1}{u} - \frac{\Lambda}{2\pi^2} - \frac{i}{4\pi} \sqrt{p^0 - \frac{\bm{p}^2}{4} - 2\,U(\bm{r}) + i\gamma_1} + O(\Lambda^{-1})\,.
\label{eq:vacuum_T_matrix_reg}
\end{equation}
To regularize $u$, we consider the on-shell T-matrix between a pair of atoms with momentum \mbox{$\bm{p}$} and \mbox{$\bm{q}$}.  Due to Galilean invariance and rotational symmetry, the $T$-matrix will only depend on the magnitude of the relative momentum between the pair of atoms $k = |\bm{p}-\bm{q}|/2$. The $s$-wave scattering length $a$ is then defined through the low-energy behavior of the $T$-matrix~\cite{BHL2017}. Working in the center of mass frame, this gives
\begin{equation}
\mathcal{T}^{-1}(k) = \lim_{|\bm{k}|\to 0}\left[T^R(X,p^0= 2\,\omega_{\bm{k}},0)\right]^{-1}  = \frac{1}{4\pi}\left(\frac{1}{a} + ik\right),
\end{equation}
where we have set $\gamma_1 = 0$ and $U = 0$. Matching this limit to Eq.~\eqref{eq:vacuum_T_matrix_reg} gives
\begin{equation}
    \frac{1}{u} = \frac{\Lambda}{2\pi^2} - \frac{1}{4\pi a}.
\end{equation}
Since $u = g + i\gamma_2$, naturally the scattering length is also complex $ a = a' - i a''$. In the main text we set $a' = 0$. In free space, the final result is the $T$-matrix determined only by the relative momentum
\begin{equation}
    \mathcal{T}(k)=\frac{4\pi}{a^{-1}+ik}\,,
\end{equation}
in agreement with the main text.

Next we consider the available phase space for scattering in Eq.~\eqref{eq:I_on_shell}. To leading order in the fugacity, we can approximate $1-f_{\sigma {\bm{p}}}\approx 1 + \mathcal{O}(z)$. This gives the final result for the kinetic theory in the dilute limit 
\begin{align}
\label{eq:I_on_shell_dilute}
I_{\sigma}^{\mathrm{in}}[f] 
&=
-\frac{\mathrm{Im}\,a^{-1}}{2\pi}\int_{\bm{q}} \left|\mathcal{T}\left(k\right)\right|^2\,f_{\sigma \bm{p}}f_{\bar{\sigma}\bm{q}}\,,\nonumber\\
I_{\sigma}^{\mathrm{el}}[f] 
&=
\int_{\bm{q}\bm{p'}\bm{q'}}\mathcal{W}_{\bm{p}\bm{q}; \bm{p'}\bm{q'}}\left[ f_{\sigma \bm{p}} f_{\bar{\sigma} \bm{q}} - f_{\sigma \bm{p'}} f_{\bar{\sigma}\bm{q'}}\right],
\end{align}
with
\begin{equation}
\mathcal{W}_{\bm{p}\bm{q}; \bm{p'}\bm{q'}}
\equiv \left|\mathcal{T}(k)\right|^2 
(2\pi)^{d+1}\,\delta^{(d)}(\bm{p}+\bm{q}-\bm{p'}-\bm{q'})\,\delta(\epsilon_{\sigma\bm{p}} + \epsilon_{\bar{\sigma}\bm{q}} - \epsilon_{\sigma\bm{p'}} -\epsilon_{\bar{\sigma}\bm{q'}})\,.
\end{equation}

\section{Chapman--Enskog expansion for lossy quantum gases}
In this section, we examine how to solve the kinetic equation presented in Eqs.~\eqref{eq:BE}\,--\,\eqref{eq:rate} in the main text. To this end we note that the system has two small parameters that characterize the transport. The first is the ratio of the absorption rate \mbox{$\Gamma^{\rm in}$} to the elastic collision rate \mbox{$\Gamma^{\rm el}$}. As discussed in the main text, for \mbox{$a'=0$} this is dictated solely by \mbox{$\Gamma^{\rm in}/\Gamma^{\rm el}\propto\lambda_T/a''$}, since the elastic and inelastic cross sections share the vertex \mbox{$|\mathcal{T}|^2$} and differ only by \mbox{$\sigma_{\rm el}/\sigma_{\rm in}=k/\mathrm{Im}\,a^{-1}=ka''$}. In the limit \mbox{$a''/\lambda_T\gg1$} the elastic collisions dominate the losses, suggesting that the distribution function can be well approximated by a time-dependent Maxwell--Boltzmann (MB) distribution up to corrections of order \mbox{$\lambda_T/a''$}. The second small parameter is the Knudsen number \mbox{$\mathrm{Kn}\sim\ell_{\rm el}/L$}, which compares the elastic mean free path \mbox{$\ell_{\rm el}=\sqrt{T}/\Gamma^{\rm el}$} with the length \mbox{$L$} over which the density \mbox{$n=2n_\sigma$} (or equivalently the fugacity \mbox{$z=n_{\sigma}\lambda_T^3$}), the velocity field \mbox{$\mathbf{u}$} and the temperature \mbox{$T$} vary. The hydrodynamic limit is characterized by \mbox{$\mathrm{Kn}\ll1$}, while the collisionless limit has \mbox{$\mathrm{Kn}\gg1$}. 

In the limit \mbox{$\mathrm{Kn}\ll1$}, we can write the distribution function as
\begin{equation}
    f_{\sigma {\bf p}} = f_{\sigma {\bf p}}^{(0)} + \delta f_{\sigma {\bf p}},
\end{equation}
where \mbox{$f_{\sigma{\bf p}}^{(0)}$} describes the slow evolution of the density, velocity field, and temperature due to the two-body losses, while \mbox{$\delta f_{\sigma{\bf p}}$} describes the relaxation of gradients in the currents. For brevity, we will suppress the explicit dependence of the distribution on the spin \mbox{$\sigma$}.

\subsection{Zeroth order}
Since the loss is slow compared with the elastic collisions, \mbox{$\Gamma^{\rm in}/\Gamma^{\rm el}\propto\lambda_T/a''\ll1$}, we can assume that the zeroth order solution \mbox{$f_{{\bf p}}^{(0)}$} is approximately given by a time-dependent MB distribution function
\begin{equation}
    f^{(0)}_{{\bf p}}({\bf r},t) = f^{\rm (LE)}_{{\bf p}} ({\bf r},t)\left[1+\psi\left(\bar{\epsilon}_{\bf p};\,\frac{a''}{\lambda_T({\bf r},t)}\right)\right],
    \label{eq:dist_zeroth_order}
\end{equation}
with
\begin{equation}
   f^{\rm (LE)}_{{\bf p}} = z(t)\exp\left(-\bar{\epsilon}_{\bf p}\right),\qquad
   \bar{\epsilon}_{\bf p} = \frac{({\bf p-u})^2}{2T(t)},
\end{equation}
being the time-dependent MB distribution, and \mbox{$\psi$} is the correction, of \mbox{$\mathcal{O}(\lambda_T/a'')$} for \mbox{$a''\gg\lambda_T$}. For \mbox{$a''\ll\lambda_T$} the vertex is momentum independent at leading order. Both the source of \mbox{$\psi$} and its elastic relaxation are then of relative order \mbox{$a''/\lambda_T$}, so the steady \mbox{$\psi$} is of order one, cf.~Fig.~\ref{fig:S1}, but it builds up only over many loss times. During the lifetime of the gas, \mbox{$\psi$} therefore remains small. Note we have explicitly written the time-dependence of the thermodynamic quantities for transparency. At zeroth order, we do not need to worry about the spatial gradients as they appear as first order corrections.

We require that \mbox{$\psi$} carries no density, momentum or energy, i.e., that the density, momentum density, and energy density computed with \mbox{$f^{(0)}_{\bf p}$} equal those of the time-dependent MB distribution
\begin{equation}
n_\sigma = \int_{\bf p} f^{\rm (LE)}_{{\bf p}},\qquad 
\mathbf{g}_\sigma = \int_{\bf p} {\bf p}\, f^{\rm (LE)}_{{\bf p}},\qquad
E_\sigma = \int_{\bf p} \epsilon_{\bf p} f^{\rm (LE)}_{{\bf p}},
\end{equation}
with \mbox{$\epsilon_{\bf p}= p^2/2$}. This implies that
\begin{align}
    0 &= \int_{\bf p} f^{\rm (LE)}_{{\bf p}}  \psi_{\bf p} =\int_{\bf p} {\bf p}\, f^{\rm (LE)}_{{\bf p}}  \psi_{\bf p} = \int_{\bf p} \epsilon_{\bf p} f^{\rm (LE)}_{{\bf p}} \psi_{\bf p}.
    \label{eq:psi_orth}
\end{align}

Inserting Eq.~\eqref{eq:dist_zeroth_order} into the Boltzmann equation, dividing by \mbox{$f^{\rm (LE)}_{\bf p}$} and keeping terms to leading order in \mbox{$\lambda_T/a''$}, one finds
\begin{equation}
\frac{\dot z}{z}+\frac{\dot T}{T}\,\bar{\epsilon}_{\mathbf{p}}+\frac{\mathbf{p-u}}{T}\cdot\dot{\mathbf{u}}=-\nu_{\mathbf{p}}-\mathcal{L}\psi_{\mathbf{p}},
\label{eq:zeroth-lin}
\end{equation}
where we have defined the loss kernel
\begin{equation}
\nu_{\bf p} = \frac{\mathrm{Im}\,a^{-1}}{2\pi}\int_{\bf q}|\mathcal{T}(k)|^2f^{\rm (LE)}_{\bf q},\qquad  k=\tfrac12|{\bf p}-{\bf q}|,
\label{eq:nu}
\end{equation}
and the linearized collision integral
\begin{equation}
\mathcal{L}\psi_{\mathbf{p}}=\int_{\mathbf{q}\mathbf{p}'\mathbf{q}'}\mathcal W\,f^{\rm (LE)}_{\mathbf{q}}\big[\psi_{\mathbf{p}}+\psi_{\mathbf{q}}-\psi_{\mathbf{p}'}-\psi_{\mathbf{q}'}\big],
\label{eq:Lop}
\end{equation}
with the transition rate \mbox{$\mathcal W$} defined in Eq.~\eqref{eq:rate}.

\subsubsection{Equations of motion for the moments}
To extract the explicit time-dependence of the density, velocity field, and temperature, we define the average \mbox{$\langle F\rangle\equiv n_\sigma^{-1}\int_{\mathbf{p}}f^{\rm (LE)}_{\mathbf{p}}F_{\mathbf{p}}$} of a function \mbox{$F_{\mathbf{p}}$} over the local MB distribution. Since \mbox{$\mathcal{L}$} conserves particle number, momentum and energy, \mbox{$\mathcal{L}\psi_{\bf p}$} drops out when projecting Eq.~\eqref{eq:zeroth-lin} on \mbox{$1$}, \mbox{$\bar{\epsilon}_{\bf p}$} and \mbox{${\bf p-u}$}, which results in the following equations of motion
\begin{equation}
\frac{\dot z}{z} =-\Big\langle\big(\tfrac52-\bar{\epsilon}_{\mathbf{p}}\big)\nu_{\mathbf{p}}\Big\rangle,\qquad
\frac{\dot T}{T} =\frac23\Big\langle\big(\tfrac32-\bar{\epsilon}_{\mathbf{p}}\big)\nu_{\mathbf{p}}\Big\rangle,\qquad
\dot{\mathbf{u}} =0.
\label{eq:zdot}
\end{equation}

Since the velocity field is unchanged by the losses, we can set \mbox{${\bf u=0}$} without loss of generality and focus on the behavior of the fugacity and the temperature. The fugacity and temperature then satisfy
\begin{equation}
\frac{\dot z}{z} = -zT\,\Lambda_z,\qquad \frac{\dot T}{T}=zT\,\Lambda_T,
\label{eq:zTode}
\end{equation}
up to corrections of order \mbox{$\lambda_T/a''$}, with
\begin{equation}
\Lambda_z = \frac{\mathrm{Im}\,a^{-1}}{(2\pi)^{5/2}}\Big[\tfrac74\big\langle|\mathcal{T}|^2\big\rangle_{\rm rel}-\tfrac12\big\langle k^2|\mathcal{T}|^2\big\rangle_{\rm rel}\Big],\qquad 
\Lambda_T = \frac{\mathrm{Im}\,a^{-1}}{(2\pi)^{5/2}}\Big[\tfrac12\big\langle|\mathcal{T}|^2\big\rangle_{\rm rel}-\tfrac13\big\langle k^2|\mathcal{T}|^2\big\rangle_{\rm rel}\Big],
\label{eq:Lambda}
\end{equation}
in units \mbox{$T=1$}. Here \mbox{$\langle g\rangle_{\rm rel}=(4/\sqrt\pi)\int_0^\infty dk\,k^2e^{-k^2}g(k)$} is the Maxwell average over the magnitude \mbox{$k$} of the relative momentum of a pair.

For \mbox{$|a|\ll\lambda_T$} the two-body $T$-matrix is constant and every momentum is lost at the same fractional rate. This preserves the Maxwellian shape of the distribution function. Then the rate equation for the total density \mbox{$n=2n_\sigma$} becomes
\begin{equation}
    \dot n(t) = -4\pi a''n^2(t),
\end{equation}
which has the solution
\begin{equation}
\frac{n(t)}{n_0} = \frac{1}{1+4\pi a'' n_0 t},\qquad
\frac{T(t)}{T_0} = \big(1+4\pi a'' n_0 t\big)^{\alpha},
\label{eq:dynweak}
\end{equation}
where \mbox{$n_0$} and \mbox{$T_0$} are the initial density and temperature, with the heating exponent
\begin{equation}
\alpha=\frac{2\sqrt2}{3}\,\frac{a''}{\lambda_T}+\pi\,\frac{|a|^2}{\lambda_T^2}+\Big(\frac{16}{3}-4\pi\Big)\frac{a''^2}{\lambda_T^2}\ll1,
\label{eq:alpha}
\end{equation}
which follows from the leading momentum dependence of the two-body $T$-matrix. It assumes a Maxwellian background, \mbox{$\psi_{\bf p}=0$}, which holds during the lifetime of the gas, see above.

For \mbox{$|a|\gg\lambda_T$} the two-body $T$-matrix reaches the unitarity limit \mbox{$|\mathcal{T}|^2=16\pi^2/k^2$}, which removes slow pairs preferentially. Then the rate equations become
\begin{equation}
\dot n =-8\pi\,\mathrm{Im}\,a^{-1}\,\frac{n^2}{T},\qquad  \frac{d\ln T}{d\ln n}=-\frac13,
\end{equation}
with solution
\begin{equation}
\frac{n}{n_0}=\Big(1+\frac{32\pi}{3} \frac{\mathrm{Im}\,a^{-1}\,n_0}{T_0}\,t\Big)^{-3/4},\qquad 
\frac{T}{T_0}=\Big(1+\frac{32\pi}{3} \frac{\mathrm{Im}\,a^{-1}\,n_0}{T_0}\,t\Big)^{1/4}.
\label{eq:dynstrong}
\end{equation}
Since \mbox{$a''/\lambda_T\propto T^{1/2}$}, \mbox{$a''/\lambda_T$} grows as the \mbox{$1/8$} power of the same factor. Corrections to Eq.~\eqref{eq:dynstrong} are of relative order \mbox{$\lambda_T/|a|$}.

We note that our result for the time dependence of the density for \mbox{$a''/\lambda_T\gg1$} differs from that of Ref.~\cite{HM2026}. The difference lies in the accessible regime. A contact coupling \mbox{$u$} with momentum cutoff \mbox{$\Lambda$} gives \mbox{$a^{-1}=4\pi/u+2\Lambda/\pi$}. The unitarity-limited regime \mbox{$a''\gg\lambda_T$} studied here requires \mbox{$\mathrm{Re}\,a^{-1}\simeq0$}, i.e., an attractive real part of \mbox{$u$} that cancels the cutoff term, as provided by a resonance with a decaying closed channel. For the purely dissipative coupling of Ref.~\cite{HM2026} there is no such cancellation, \mbox{$\mathrm{Re}\,a^{-1}=2\Lambda/\pi$}, and \mbox{$|a|\le\pi/(2\Lambda)$} for any loss rate. With \mbox{$\Lambda$} set by the lattice spacing, the gas remains in the regime \mbox{$|a|\ll\lambda_T$}, where the $T$-matrix is constant and we recover their result, \mbox{$n\propto t^{-1}$} at constant temperature.

\subsubsection{Deviation due to dissipation}

We now evaluate the deviation \mbox{$\psi_{\bf p}$} of the distribution function due to the losses. Equation~\eqref{eq:zeroth-lin} is sufficient for the moments, since the remaining terms linear in \mbox{$\psi_{\bf p}$} give corrections of order \mbox{$b_n$}, i.e., \mbox{$\lambda_T/a''$} for \mbox{$a''\gg\lambda_T$}, and relative corrections of order \mbox{$b_n$} to the heating exponent for \mbox{$a''\ll\lambda_T$}. For \mbox{$\psi_{\bf p}$} itself we keep all terms linear in \mbox{$\psi_{\bf p}$} in the zeroth order Boltzmann equation
\begin{equation}
\mathcal{L}\psi_{\bf p}+\delta\nu_{\bf p}\,\psi_{\bf p}+\mathcal{J}\psi_{\bf p}=-\delta\nu_{\bf p},\qquad
\delta\nu_{\bf p}=\nu_{\bf p}+\frac{\dot z}{z}+\frac{\dot T}{T}\,\bar{\epsilon}_{\bf p}.
\label{eq:psifull}
\end{equation}
Here \mbox{$\mathcal{L}\psi_{\bf p}$} is the elastic scattering term. The term \mbox{$\delta\nu_{\bf p}\psi_{\bf p}$} combines the loss of the distorted particles at their own rate, \mbox{$\nu_{\bf p}\psi_{\bf p}$}, with the time derivative of \mbox{$f^{\rm (LE)}_{\bf p}$} acting on \mbox{$\psi_{\bf p}$}, \mbox{$(\dot z/z+\bar{\epsilon}_{\bf p}\dot T/T)\psi_{\bf p}$}. The final term \mbox{$\mathcal{J}\psi_{\bf p}$} describes the loss of their collision partners
\begin{equation}
\mathcal{J}\psi_{\bf p}=\frac{\mathrm{Im}\,a^{-1}}{2\pi}\int_{\bf q}|\mathcal{T}(k)|^2f^{\rm (LE)}_{\bf q}\,\psi_{\bf q},
\label{eq:Jop}
\end{equation}
with \mbox{$\bf k = (p-q)/2$} being the relative momentum of the pair. All terms are proportional to \mbox{$zT$}, so \mbox{$\psi_{\bf p}$} is independent of the density and depends on time only through \mbox{$a/\lambda_T(t)$}. We neglect \mbox{$\dot\psi_{\bf p}$}. For \mbox{$a''\gg\lambda_T$} this is justified because elastic collisions relax \mbox{$\psi_{\bf p}$} much faster than the gas decays. For \mbox{$a''\ll\lambda_T$} it instead yields the steady-state distortion, which a gas prepared in equilibrium does not reach during its lifetime. We note that for \mbox{$a'' = 0$}, i.e.\ for real scattering lengths, \mbox{$\delta \nu_{\bf p} = \mathcal{J} = 0$}. Therefore one obtains the solution \mbox{$\psi_{\bf p}=0$}, as expected from the Chapman--Enskog expansion in the absence of losses. 

The distortion \mbox{$\psi_{\bf p}$} depends only on \mbox{$\bar{\epsilon}_{\bf p}$} and must satisfy Eq.~\eqref{eq:psi_orth}. We therefore expand \mbox{$\psi_{\bf p}$} in the generalized Laguerre (Sonine) polynomials \mbox{$L^{(1/2)}_n(\bar{\epsilon})$}. These are orthogonal with respect to the MB distribution,
\begin{equation}
\int_{\bf p}f_{\bf p}^{\rm (LE)}L_{n,{\bf p}}L_{m,{\bf p}} = N_{n} \delta_{n,m},\qquad L_{n,{\bf p}} = L^{(1/2)}_n(\bar{\epsilon}_{\bf p}),
\end{equation}
with the normalization \mbox{$N_n=n_\sigma\Gamma(n+\frac32)/[n!\,\Gamma(\frac32)]$}.
In terms of these polynomials, \mbox{$\psi_{\bf p}$} can be written as
\begin{equation}
\psi_{\bf p}=\sum_{n\ge2}b_n\, L_{n,{\bf p}},
\label{eq:sonine}
\end{equation}
with Sonine coefficients \mbox{$b_n$}. The lowest two polynomials, 
\mbox{$L^{(1/2)}_0(\bar{\epsilon})=1$} and \mbox{$L^{(1/2)}_1(\bar{\epsilon})=\frac32-\bar{\epsilon}$} are excluded from the sum as they span the sectors \mbox{$\{1,\bar{\epsilon}\}$}. 

Multiplying Eq.~\eqref{eq:psifull} by \mbox{$f^{\rm (LE)}_{\bf p}L_{m,{\bf p}}$} with \mbox{$m\ge2$} and integrating over \mbox{${\bf p}$} gives a linear system of equations for the expansion coefficients \mbox{$b_n$}
\begin{equation}
\sum_{n\ge2}b_n\int_{\bf p}f^{\rm (LE)}_{\bf p}L_{m,{\bf p}}\left[\mathcal{L}L_{n,{\bf p}}+\delta\nu_{\bf p}L_{n,{\bf p}}+\mathcal{J}L_{n,{\bf p}}\right]= -\int_{\bf p}f^{\rm (LE)}_{\bf p}L_{m,{\bf p}}\,\nu_{\bf p},
\label{eq:galerkin}
\end{equation}
where \mbox{$\mathcal{L}L_{n,{\bf p}}$} and \mbox{$\mathcal{J}L_{n,{\bf p}}$} denote Eqs.~\eqref{eq:Lop} and \eqref{eq:Jop} evaluated with \mbox{$\psi_{\bf p}=L_{n,{\bf p}}$}. Only \mbox{$\nu_{\bf p}$} contributes to the right side, because \mbox{$L_{m,{\bf p}}$} with \mbox{$m\ge2$} is orthogonal to \mbox{$1$} and \mbox{$\bar{\epsilon}_{\bf p}$}.

\begin{figure*}
    \centering
    \includegraphics[width=\linewidth]{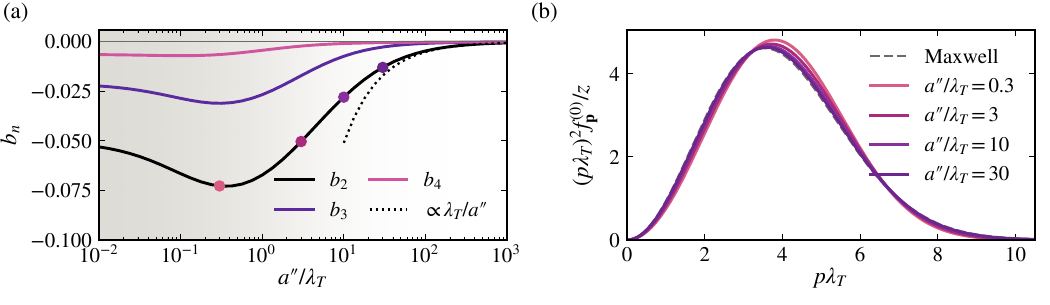}
    \caption{(a) Sonine coefficients \mbox{$b_n$} obtained from Eq.~\eqref{eq:galerkin} as a function of \mbox{$a''/\lambda_T$}. Dotted, the asymptote \mbox{$b_2\simeq\beta_2\lambda_T/a''$} of Eq.~\eqref{eq:binf}. Shading: as in Fig.~\ref{fig:2}. The colored dots along \mbox{$b_2$} mark the values of \mbox{$a''/\lambda_T$} used in (b). (b) Radial momentum distribution \mbox{$(p\lambda_T)^2f^{(0)}_{\bf p}/z$} at fixed \mbox{$a''/\lambda_T$} (solid lines), compared with the instantaneous local equilibrium (dashed line).}
    \label{fig:S1}
\end{figure*}

We solve Eq.~\eqref{eq:galerkin} by truncating the system at \mbox{$n=6$}. Figure~\ref{fig:S1}(a) collects the Sonine coefficients up to \mbox{$n=4$}. The magnitude of the Sonine coefficients is largest near \mbox{$a''\approx0.3\lambda_T$}, tends to a constant for \mbox{$a''\ll\lambda_T$}, and decays as \mbox{$\lambda_T/a''$} for \mbox{$a''/\lambda_T\gg 1$}. In the limit \mbox{$a''/\lambda_T\gg 1$} one can obtain the asymptotic values of these expansion coefficients
\begin{equation}
b_n\simeq\beta_n\,\frac{\lambda_T}{a''},\qquad 
\beta_2=-0.509,\qquad  \beta_3=-0.075,\qquad \beta_4=-0.020,
\label{eq:binf}
\end{equation}
as shown for \mbox{$b_2$} by the dotted line. Fig.~\ref{fig:S1}(b) shows the resulting distribution function at fixed instantaneous \mbox{$a''/\lambda_T$}. The deviation is largest for \mbox{$a''\lesssim\lambda_T$}, and it depletes low momenta and shifts the maximum to higher momentum.

\subsection{First order}
 
We insert \mbox{$f_{\bf p}=f^{(0)}_{\bf p}+\delta f_{\bf p}$} into the Boltzmann equation and keep terms of first order in \mbox{$\mathrm{Kn}$}. The terms of zeroth order cancel by Eqs.~\eqref{eq:zeroth-lin} and \eqref{eq:psifull}, and the remainder is
\begin{equation}
\partial_t\,\delta f_{\bf p}-I'_{\bf p}[\delta f]=-f^{(0)}_{\bf p}H_{\bf p},
\label{eq:first}
\end{equation}
where \mbox{$I'_{\bf p}$} is the collision integral linearized about \mbox{$f^{(0)}_{\bf p}$}, and \mbox{$f^{(0)}_{\bf p}H_{\bf p}$} collects all terms of first order in \mbox{$\mathrm{Kn}$} that involve \mbox{$f^{(0)}_{\bf p}$} alone.
 
The left side of Eq.~\eqref{eq:first} describes the dynamics of the deviation. The term \mbox{$\partial_t\,\delta f_{\bf p}$} does not vanish once the fluxes have relaxed, because the background decays and heats at the loss rate. The term \mbox{$I'_{\bf p}[\delta f]$} relaxes the deviation through elastic collisions and losses. The right side is the source and describes the effects of spatially inhomogeneous density, temperature and velocity fields.
 
First consider the left hand side of Eq.~\eqref{eq:first}. Linearizing the collision integrals of Eqs.~\eqref{eq:Elastic} and \eqref{eq:Inelastic} of the main text about \mbox{$f^{(0)}_{\bf p}$} gives \mbox{$I'_{\bf p}=I'^{\,\rm el}_{\bf p}+I'^{\,\rm in}_{\bf p}$}, with
\begin{equation}
I'^{\,\rm el}_{\bf p}[\delta f]=\int_{{\bf q}{\bf p}'{\bf q}'}\mathcal W\big[f^{(0)}_{{\bf p}'}\delta f_{{\bf q}'}+\delta f_{{\bf p}'}f^{(0)}_{{\bf q}'}-f^{(0)}_{\bf p}\delta f_{\bf q}-\delta f_{\bf p}f^{(0)}_{\bf q}\big],\qquad
I'^{\,\rm in}_{\bf p}[\delta f]=-\nu_{\bf p}\,\delta f_{\bf p}-\frac{\mathrm{Im}\,a^{-1}}{2\pi}\,f^{(0)}_{\bf p}\int_{\bf q}|\mathcal{T}(k)|^2\,\delta f_{\bf q},
\label{eq:Ilin}
\end{equation}
where \mbox{${\bf q}$} is the momentum of the collision partner, \mbox{${\bf p}'$} and \mbox{${\bf q}'$} are the outgoing momenta, and \mbox{$k$} is defined in Eq.~\eqref{eq:nu}. The elastic part describes scattering of the deviation off partners drawn from \mbox{$f^{(0)}$}. Since \mbox{$f^{(0)}_{\bf p}$} is not a MB distribution, it does not satisfy detailed balance, \mbox{$f^{(0)}_{{\bf p}'}f^{(0)}_{{\bf q}'}\neq f^{(0)}_{\bf p}f^{(0)}_{\bf q}$}, and the distortion \mbox{$\psi_{\bf p}$} modifies the elastic relaxation of the deviation. The first term of the loss part in Eq.~\eqref{eq:Ilin} describes the loss of the particle that carries the deviation, at the rate \mbox{$\nu_{\bf p}$} [
cf.\ Eq.~\eqref{eq:nu}]. Here we have taken the partners from \mbox{$f^{\rm (LE)}_{\bf q}$} instead of \mbox{$f^{(0)}_{\bf q}$}, which changes \mbox{$\nu_{\bf p}$} only at higher order in \mbox{$\lambda_T/a''$}. The second part of the linearized loss collision integral is the loss of the collision partners. The elastic part conserves particle number, momentum and energy. The loss part does not. In particular it removes momentum from the deviation at the rate
\begin{equation}
{\bf F}[\delta f]=\int_{\bf p}{\bf p}\,I'^{\,\rm in}_{\bf p}[\delta f].
\label{eq:force}
\end{equation}
 
The right hand side of Eq.~\eqref{eq:first} describes the effects of the time derivatives and spatial gradients of \mbox{$z$}, \mbox{$T$} and \mbox{${\bf u}$}, through which \mbox{$f^{(0)}_{\bf p}$} depends on position and time. The time derivative of \mbox{$f^{(0)}_{\bf p}$} consists of the loss-driven part, with the rates \mbox{$\dot z$}, \mbox{$\dot T$} and \mbox{$\dot{\bf u}$} obtained from Eq.~\eqref{eq:zdot}, and a gradient-driven part. The source in Eq.~\eqref{eq:first} is the streaming term together with the gradient-driven part,
\begin{equation}
H_{\bf p}=\Big({\bf p}\cdot\nabla_{\bf R}-\nabla U_{\rm ext}\cdot\nabla_{\bf p}\Big)\ln f^{(0)}_{\bf p}+\frac{\partial\ln f^{(0)}_{\bf p}}{\partial\ln z}\Big(\partial_t\ln z-\frac{\dot z}{z}\Big)+\frac{\partial\ln f^{(0)}_{\bf p}}{\partial\ln T}\Big(\partial_t\ln T-\frac{\dot T}{T}\Big)+\frac{\partial\ln f^{(0)}_{\bf p}}{\partial{\bf u}}\cdot\partial_t{\bf u},
\label{eq:Hdef}
\end{equation}
where \mbox{$U_{\rm ext}$} is an external trapping potential, set to zero in Eq.~\eqref{eq:BE} of the main text. To evaluate Eq.~\eqref{eq:Hdef} we need the time derivatives of \mbox{$z$}, \mbox{$T$} and \mbox{${\bf u}$} to first order in \mbox{$\mathrm{Kn}$}. They follow from the moments of the Boltzmann equation with \mbox{$1$}, \mbox{${\bf p}$} and \mbox{$\epsilon_{\bf p}$}, evaluated with \mbox{$f^{(0)}_{\bf p}$}. The momentum flux of \mbox{$f^{(0)}_{\bf p}$} is the pressure \mbox{$P=nT$}, exactly, because \mbox{$\psi_{\bf p}$} carries neither density nor energy, and its heat current vanishes. The moments are therefore the Euler equations, supplemented by the loss,
\begin{align}
\partial_t n+\nabla\cdot(n{\bf u})&=\dot n,\nonumber\\
\partial_t T+{\bf u}\cdot\nabla T+\tfrac23T\,\nabla\cdot{\bf u}&=\dot T,\nonumber\\
\partial_t{\bf u}+({\bf u}\cdot\nabla){\bf u}&=-\frac{\nabla P}{n}-\nabla U_{\rm ext}+\frac{{\bf F}[\delta f]}{n_\sigma},
\label{eq:hydro}
\end{align}
where \mbox{$\dot n$} and \mbox{$\dot T$} are the loss rates of Eq.~\eqref{eq:zdot} evaluated with the local density and temperature. The force \mbox{${\bf F}$} is the momentum removed by the loss, Eq.~\eqref{eq:force}. Equations~\eqref{eq:hydro} are the solvability conditions of Eq.~\eqref{eq:first}. With them, the deviation carries no density, momentum or energy,
\begin{equation}
0=\int_{\bf p}\delta f_{\bf p}=\int_{\bf p}{\bf p}\,\delta f_{\bf p}=\int_{\bf p}\epsilon_{\bf p}\,\delta f_{\bf p}.
\label{eq:matching}
\end{equation}
 
We evaluate the source at a point where \mbox{${\bf u}=0$}. Since \mbox{$n\propto zT^{3/2}$} and \mbox{$\nabla P/(nT)=\nabla\ln z+\frac52\nabla\ln T$}, Eqs.~\eqref{eq:hydro} give there
\begin{align}
\partial_t\ln z-\frac{\dot z}{z}&=0,\nonumber\\
\partial_t\ln T-\frac{\dot T}{T}&=-\tfrac23\nabla\cdot{\bf u},\nonumber\\
\partial_t{\bf u}&=-T\Big(\nabla\ln z+\tfrac52\nabla\ln T\Big)-\nabla U_{\rm ext}+\frac{{\bf F}[\delta f]}{n_\sigma}.
\label{eq:dtX}
\end{align}
The derivatives of \mbox{$f^{(0)}_{\bf p}$}, Eq.~\eqref{eq:dist_zeroth_order}, are taken to first order in \mbox{$\psi_{\bf p}$}. They involve the energy derivative \mbox{$\partial_{\bar{\epsilon}}\psi_{\bf p}\equiv d\psi/d\bar{\epsilon}$} at \mbox{$\bar{\epsilon}=\bar{\epsilon}_{\bf p}$}, and the temperature derivative \mbox{$\chi_{\bf p}\equiv\partial\psi_{\bf p}/\partial\ln T$} at fixed \mbox{$\bar{\epsilon}_{\bf p}$} and \mbox{$a$}, which arises because \mbox{$\psi_{\bf p}$} depends on \mbox{$a''/\lambda_T$}.

Inserting Eq.~\eqref{eq:dtX} and these derivatives into Eq.~\eqref{eq:Hdef}, and using the traceless strain rate \mbox{$V_{ij}=\partial_iu_j+\partial_ju_i-\frac23\delta_{ij}\nabla\cdot{\bf u}$}, the source decomposes into four parts, \mbox{$H_{\bf p}=H^\eta_{\bf p}+H^\kappa_{\bf p}+H^{0}_{\bf p}+H^F_{\bf p}$},
where 
\begin{subequations}
\begin{align}
H^\eta_{\bf p}&=\frac{1}{2T}\big(1-\partial_{\bar{\epsilon}}\psi_{\bf p}\big)\,p_ip_jV_{ij},\label{eq:shear_strain}\\
H^\kappa_{\bf p}&=\Big[\big(1-\partial_{\bar{\epsilon}}\psi_{\bf p}\big)\big(\bar{\epsilon}_{\bf p}-\tfrac52\big)+\chi_{\bf p}\Big]\,{\bf p}\cdot\nabla\ln T+\partial_{\bar{\epsilon}}\psi_{\bf p}\,{\bf p}\cdot\nabla\ln z,\label{eq:thermal_strain} \\
H^{0}_{\bf p}&=-\tfrac23\chi_{\bf p}\,\nabla\cdot{\bf u}, \label{eq:bulk_strain}
\end{align}
\end{subequations}
 
Equations~\eqref{eq:shear_strain}\,--\,\eqref{eq:bulk_strain} describe
 the change in the distribution function due to shear strain, temperature and density gradients, bulk strain, respectively. There can also be a force like strain due to the loss, however we neglect this as it doesn't affect the shear viscosity and thermal conductivity.For a MB distribution, \mbox{$\psi_{\bf p}=0$}, these are the familiar sources \mbox{$p_ip_jV_{ij}/2T$} and \mbox{$(\bar{\epsilon}_{\bf p}-\frac52)\,{\bf p}\cdot\nabla\ln T$}. The distortion modifies both and adds a response to density gradients at fixed temperature, the last term of \mbox{$H^\kappa_{\bf p}$}. The scalar part \mbox{$H^{0}_{\bf p}$} is isotropic and contributes neither to the shear stress nor to the energy current, and we do not consider it further. The force part \mbox{$H^F_{\bf p}$} is linear in \mbox{$\delta f$}, and we move it to the left side of Eq.~\eqref{eq:first}, which becomes
\begin{equation}
\partial_t\,\delta f_{\bf p}-I'_{\bf p}[\delta f]+\frac{f^{(0)}_{\bf p}}{n_\sigma T}\,{\bf p}\cdot{\bf F}[\delta f]=-f^{(0)}_{\bf p}\big(H^\eta_{\bf p}+H^\kappa_{\bf p}\big).
\label{eq:first-final}
\end{equation}
 
\subsection{Mode expansion at first order}
In order to analyse Eq.~\eqref{eq:first-final} we note that the collision integral and the force term in Eq.~\eqref{eq:first-final} do not mix deviations of different angular symmetry. Hence we can treat the shear and heat sources independently. We take a strain rate with the single component \mbox{$V_{xy}=V_{yx}$} and gradients along \mbox{$x$}, and write the sources of Eqs.~\eqref{eq:shear_strain} and \eqref{eq:thermal_strain} as a thermodynamic force \mbox{$X^c$} times a function of momentum \mbox{$h^c_{\bf p}$},
\begin{equation}
H^\eta_{\bf p}=X^\eta h^\eta_{\bf p},\qquad 
H^\kappa_{\bf p}=X^\kappa h^\kappa_{\bf p}+X^z h^z_{\bf p},
\label{eq:Hsplit}
\end{equation}
with
\begin{align}
X^\eta&=\frac{V_{xy}}{T}, & h^\eta_{\bf p}&=\big(1-\partial_{\bar{\epsilon}}\psi_{\bf p}\big)\,p_xp_y,\nonumber\\
X^\kappa&=\partial_x\ln T, & h^\kappa_{\bf p}&=\Big[\big(1-\partial_{\bar{\epsilon}}\psi_{\bf p}\big)\big(\bar{\epsilon}_{\bf p}-\tfrac52\big)+\chi_{\bf p}\Big]p_x,\nonumber\\
X^z&=\partial_x\ln z, & h^z_{\bf p}&=\partial_{\bar{\epsilon}}\psi_{\bf p}\,p_x.
\label{eq:forces}
\end{align}
 
Each force \mbox{$X^c$}, \mbox{$c=\eta,\kappa,z$}, drives a deviation \mbox{$\delta f^c_{\bf p}$}. We then expand \mbox{$\delta f^c_{\bf p}$}
in terms of an orthogonal set of basis functions \mbox{$U^c_{j,{\bf p}}$} with amplitudes \mbox{$d^c_j$} \cite{Hofmann2023,MakiEnss2023}
\begin{align}
\delta f^c_{\bf p}&=-X^c f^{(0)}_{\bf p}\sum_{j=1}^{N}d^c_j\,U^c_{j,{\bf p}}.
\label{eq:modes}
\end{align}
The basis functions are \mbox{$p_xp_y$} for shear and \mbox{$p_x$} for heat, each times a polynomial in \mbox{$\bar{\epsilon}_{\bf p}$}. The heat basis is shared by \mbox{$c=\kappa$} and \mbox{$c=z$}. The functions are orthogonal with respect to the MB distribution, \mbox{$\int_{\bf p}f^{\rm (LE)}_{\bf p}U^c_{i,{\bf p}}U^c_{j,{\bf p}}=N^c_i\,\delta_{ij}$}. The first basis function is chosen so that its moment is the flux, \mbox{$U^\eta_{1,{\bf p}}=p_xp_y$} and \mbox{$U^\kappa_{1,{\bf p}}=p_x(\bar{\epsilon}_{\bf p}-\frac52)$}. The heat basis is in addition orthogonal to \mbox{$p_x$} with respect to \mbox{$f^{\rm (LE)}$}. Since the deviation is weighted with \mbox{$f^{(0)}$}, it carries momentum \mbox{$\int_{\bf p}p_x\,\delta f^\kappa_{\bf p}\propto b_2N^\kappa_1$}. Matching, Eq.~\eqref{eq:matching}, is restored by \mbox{$U^\kappa_{1,{\bf p}}\to U^\kappa_{1,{\bf p}}-\frac52b_2\,p_x$}, which leaves the weights \mbox{$W^c_j$} below unchanged at linear order in \mbox{$b_n$}.
We note that higher basis functions can be obtained using  the Gram--Schmidt process \cite{Hofmann2023}.
 
Once the fluxes have relaxed, the term \mbox{$\partial_t\,\delta f_{\bf p}$} can be dropped. Inserting Eq.~\eqref{eq:modes} into Eq.~\eqref{eq:first-final}, multiplying by \mbox{$U^c_{i,{\bf p}}$}, and integrating over \mbox{${\bf p}$} then gives a linear system for the amplitudes,
\begin{align}
\sum_{j=1}^{N}A^c_{ij}\,d^c_j&=r^c_i,
\label{eq:Agamma}
\end{align}
with
\begin{equation}
A^c_{ij}=-\frac{1}{N^c_1}\int_{\bf p}U^c_{i,{\bf p}}\,I'_{\bf p}\big[f^{(0)}U^c_j\big],\qquad 
r^c_i=\frac{1}{N^c_1}\int_{\bf p}f^{(0)}_{\bf p}U^c_{i,{\bf p}}\,h^c_{\bf p}.
\label{eq:Ar}
\end{equation}
The force term of Eq.~\eqref{eq:first-final} vanishes in the shear channel. In the heat channel it enters only through the momentum of order \mbox{$b_2$} above, i.e., at second order in \mbox{$\lambda_T/a''$} for \mbox{$a''\gg\lambda_T$}, and we drop it. The collision matrix consists of an elastic part, from \mbox{$I'^{\,\rm el}_{\bf p}$}, and a loss part, from \mbox{$I'^{\,\rm in}_{\bf p}$}. In the loss part we take the partners and the loss rate from \mbox{$f^{\rm (LE)}_{\bf p}$} and keep the distortion only in the overlap of the basis functions,
\begin{equation}
A^{c,{\rm in}}_{ij}=\sum_{k=1}^{N}G^c_{ik}\,\frac{N^c_1}{N^c_k}\,\bar A^{c,{\rm in}}_{kj},\qquad
G^c_{ik}=\frac{1}{N^c_1}\int_{\bf p}f^{(0)}_{\bf p}U^c_{i,{\bf p}}U^c_{k,{\bf p}},
\label{eq:Ain}
\end{equation}
where \mbox{$\bar A^{c,{\rm in}}_{kj}$} is the loss part of Eq.~\eqref{eq:Ar} with \mbox{$f^{(0)}$} replaced by \mbox{$f^{\rm (LE)}$}. This is exact for a momentum-independent $T$-matrix and differs from the full projection only at second order in \mbox{$\lambda_T/a''$} for \mbox{$a''\gg\lambda_T$}. For \mbox{$a''\ll\lambda_T$} it neglects relative corrections of order \mbox{$b_n$}, which do not build up during the lifetime of the gas, cf.~Sec.~\ref{sec:weak}.
 
Once we have evaluated the deviation of the distribution function, we calculate the corrections to the off-diagonal elements of the stress tensor and the energy current
\begin{equation}
\Pi_{xy}=\sum_\sigma\int_{\bf p} p_x p_y\, \delta f_{\bf p} = -\eta\,V_{xy},\qquad
j_x^{\epsilon}=\sum_\sigma\int_{\bf p} \epsilon_{\bf p}\, p_x\, \delta f_{\bf p}=-\kappa\,\partial_xT-\tfrac52nT^2\tau_z\,\partial_x\ln z,
\label{eq:fluxes}
\end{equation}
with the shear viscosity \mbox{$\eta$}, the thermal conductivity \mbox{$\kappa$} and the Prandtl number \mbox{$\mathrm{Pr}$},
\begin{equation}
\eta=nT\,\tau_\eta,\qquad 
\kappa=\tfrac52nT\,r_T\,\tau_\kappa,\qquad
\mathrm{Pr}=\frac{c_p\,\eta}{\kappa}=\frac{\tau_\eta}{r_T\,\tau_\kappa},
\label{eq:transport}
\end{equation}
where \mbox{$c_p=\frac52$} is the heat capacity per particle at constant pressure and \mbox{$r_T\equiv r^\kappa_1$} is the drive of the energy current by a temperature gradient. The relaxation times follow from the projections of the amplitudes on the flux,
\begin{equation}
\tau_\eta=\sum_{j=1}^{N}W^\eta_j\,d^\eta_j,\qquad
r_T\,\tau_\kappa=\sum_{j=1}^{N}W^\kappa_j\,d^\kappa_j,\qquad
\tau_z=\sum_{j=1}^{N}W^\kappa_j\,d^z_j,
\label{eq:tau}
\end{equation}
with the weights \mbox{$W^c_j=G^c_{1j}$}.
 
\subsubsection{Relaxation-time approximation}
\label{sec:RTA}
 
In the relaxation-time approximation we keep a single mode, \mbox{$N=1$}. The drives and weights are
\begin{align}
r^\eta_1&=1, & r^\kappa_1&=r_T=1+\tfrac72b_2+c_2, & r^z_1&=r_z=b_2,\nonumber\\
W^\eta_1&=1+b_2, & W^\kappa_1&=1+\tfrac{11}{2}b_2-\tfrac72b_3,
\label{eq:rW}
\end{align}
with \mbox{$c_2=\partial b_2/\partial\ln T$} at fixed \mbox{$a$}, the \mbox{$L_2$} component of \mbox{$\chi_{\bf p}$}. The distortion reduces the drive of the heat current by a temperature gradient, and a density gradient drives a heat current through \mbox{$r_z$}. Equations~\eqref{eq:Agamma} and \eqref{eq:tau} then give
\begin{equation}
\tau_\eta=\frac{W^\eta_1}{A^\eta_{11}},\qquad
\tau_\kappa=\frac{W^\kappa_1}{A^\kappa_{11}},\qquad 
\tau_z=r_z\,\tau_\kappa.
\label{eq:tauRTA}
\end{equation}
 
With a single mode the time derivative can be kept. The moments of Eq.~\eqref{eq:first-final} with \mbox{$p_xp_y$} and \mbox{$p_x(\epsilon_{\bf p}-\frac52T)$} close, because the collision moment of a single mode is proportional to the flux it carries. The time dependence of the weight \mbox{$\frac52T$} does not contribute, by Eq.~\eqref{eq:matching}. The stress and the energy current then obey relaxation equations,
\begin{subequations}
\begin{align}
\tau_{\eta}\,\partial_t\Pi_{xy}+\Pi_{xy}&=-\eta\,V_{xy},
\label{eq:shear_pert_SM}\\
\tau_\kappa\,\partial_t j_x^{\epsilon}+j_x^{\epsilon}&=-\kappa\,\partial_xT-\tfrac52nT^2r_z\,\tau_\kappa\,\partial_x\ln z.
\label{eq:thermal_pert_SM}
\end{align}    
\end{subequations}
These relaxation equations define \mbox{$\tau_\eta$} and \mbox{$\tau_\kappa$}, whose limits are Eqs.~\eqref{eq:tau_strong} and \eqref{eq:tau_weak} of the main text. Alternatively, the fluxes can follow their Navier--Stokes values adiabatically, \mbox{$\Pi_{xy}=-\eta^{\rm NS}V_{xy}$}. Equation~\eqref{eq:shear_pert_SM} then gives \mbox{$\eta^{\rm NS}=\eta/(1+\tau_\eta\,\partial_t\ln\eta^{\rm NS})$}, and Eq.~\eqref{eq:thermal_pert_SM} gives the analogous coupled result for \mbox{$\kappa^{\rm NS}$} and \mbox{$\tau_z$}. Since the background changes at the loss rate, the two forms differ at order \mbox{$\tau\Gamma^{\rm in}\sim\lambda_T/a''$}. This is the order of the loss corrections in Eq.~\eqref{eq:tau_strong}, which therefore depend on how the relaxation of the fluxes is defined. We quote them in the relaxation form, consistent with \mbox{$\eta=nT\tau_\eta$}.
 
\subsubsection{Loss-dominated limit}
\label{sec:weak}

For \mbox{$a''\ll\lambda_T$}, the relaxation rates \mbox{$1/\tau_{\eta(\kappa)}$} are dominated by the loss terms of Eq.~\eqref{eq:Ar}. To leading order the $T$-matrix is momentum independent, and both fluxes decay at the mean loss rate \mbox{$\langle\nu\rangle=4zT\,a''/\lambda_T$}. The next order follows from \mbox{$|\mathcal{T}|^2\propto a''^2/(1+ka'')^2\simeq a''^2(1-2ka'')$}, with \mbox{$k=|{\bf p}-{\bf l}|/2$} the relative momentum of the pair. Pairs with large \mbox{$k$} are lost less often. A flux \mbox{$U_{\bf p}$} weights fast particles, so it decays more slowly than the density. Including the simultaneous loss of the partner \mbox{${\bf l}$}, the loss part of the rate reads
\begin{equation}
\frac{\Gamma^{\rm in}_{\eta(\kappa)}}{zT}=4\frac{a''}{\lambda_T}-16\sqrt2\,R_{\eta(\kappa)}\frac{a''^2}{\lambda_T^2},\qquad R_c=\frac{\langle U_{\bf p}^2k\rangle+\langle U_{\bf p}U_{\bf l}k\rangle}{\langle U_{\bf p}^2\rangle\langle k\rangle},
\label{eq:Gin_weak}
\end{equation}
where \mbox{$\langle\cdot\rangle$} averages over two Maxwellians. With \mbox{$U^\eta_{\bf p}=p_xp_y$} and \mbox{$U^\kappa_{\bf p}=p_x(\epsilon_{\bf p}-\frac52T)$}, the Gaussian integrals give \mbox{$R_\eta=79/60-1/60=13/10$} and \mbox{$R_\kappa=341/240-3/240=169/120$}, where the second terms stem from the partner. The elastic collisions add the Born rate \mbox{$(32\sqrt2/5d_{\eta(\kappa)})\,zT\,a''^2/\lambda_T^2$}, with \mbox{$d_{\eta(\kappa)}$} as in Eq.~\eqref{eq:tau_strong}. Inverting the total rate yields Eq.~\eqref{eq:tau_weak} with
\begin{equation}
C_{\eta(\kappa)}=\sqrt2\left(R_{\eta(\kappa)}-\frac{2}{5d_{\eta(\kappa)}}\right),
\label{eq:C_weak}
\end{equation}
that is, \mbox{$C_\eta=9\sqrt2/10$} and \mbox{$C_\kappa=137\sqrt2/120$}, in agreement with the numerical solution for a Maxwellian background. A steady distortion \mbox{$\psi$} would shift \mbox{$C$} at the same order, but it does not build up during the lifetime of the gas.

\subsubsection{Strong-dissipation limit}

For \mbox{$a''\gg\lambda_T$} the T-matrix of Eq.~(3) approaches its unitary value \mbox{$|\mathcal{T}(\boldsymbol{k})|^2=16\pi^2/|\boldsymbol{k}|^2$}, and Eq.~(S106) reduces to the relaxation times of the unitary Fermi gas, \mbox{$\tau_{\eta(\kappa)}=d_{\eta(\kappa)}\tau_u/(zT)$}, with the unitary constant \mbox{$\tau_u=15\pi/(16\sqrt2)$} and \mbox{$d_{\eta(\kappa)}=1\,(3/2)$}. To first order in \mbox{$\lambda_T/a''$}, one recovers Eq.~\eqref{eq:tau_strong} of the main text
\begin{align}
\tau_{\eta(\kappa)}&=\frac{d_{\eta(\kappa)}\tau_u}{zT}\left(1-c_{\eta(\kappa)}\frac{\lambda_T}{a''}\right),
\end{align}
with
\begin{equation}
C_\eta =\frac{3}{2\sqrt2}+0.148=1.208,\qquad 
C_\kappa=\frac{27}{8\sqrt2}+0.577=2.964,
\end{equation}
where the second terms stem from the distortion. In the same limit \mbox{$r_T=1+3b_2=1-1.53\,\lambda_T/a''$}, so that the thermal conductivity \mbox{$\kappa=\frac52nTr_T\tau_\kappa$} carries the combined correction \mbox{$1-4.49\,\lambda_T/a''$}.

\subsubsection{Two modes}

Figure~\ref{fig:S2} compares the previous result to the relaxation time calculated using two modes. For the elastic gas at unitarity the ratio of the two is \mbox{$193/190$} for shear and \mbox{$41/40$} for heat. For \mbox{$a''\gg\lambda_T$} the change approaches these unitary values. It is largest near \mbox{$a''/\lambda_T\simeq2$}--\mbox{$5$}, where it reaches \mbox{$3.4\%$} for shear and \mbox{$5.7\%$} for heat. In the loss-dominated limit the ratio approaches one, and with a Maxwellian background a single mode reproduces the constants \mbox{$C_{\eta(\kappa)}$} of Eq.~\eqref{eq:C_weak}.
 
\begin{figure*}
\centering
\includegraphics[width=\linewidth]{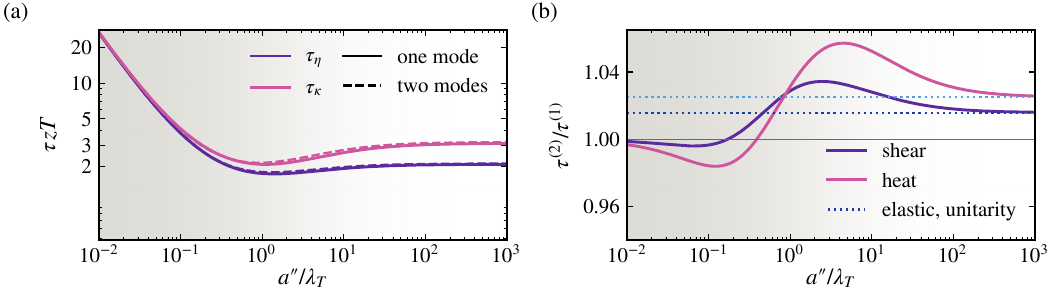}
\caption{(a) Relaxation times \mbox{$\tau_\eta$} (purple) and \mbox{$\tau_\kappa$} (magenta) in units of \mbox{$1/(zT)$} as calculated using one (solid) and two modes (dashed). (b) Ratio of the two-mode relaxation time to the one-mode result. Dotted lines represent the elastic values at unitarity. Shading: as in Fig.~\ref{fig:2}.}
\label{fig:S2}
\end{figure*}

\end{widetext}

\bibliography{Bibliography}

\end{document}